# The jigsaw puzzle of sequence phenotype inference: Piecing together Shannon entropy, importance sampling, and Empirical Bayes

Zeina Shreif[1]
shreifzz@mail.nih.gov

Deborah A. Striegel[1]
Deborah.striegel@nih.gov

Vipul Periwal[1*]
vipulp@mail.nih.gov

---

[1] Laboratory of Biological Modeling, National Institutes of Diabetes and Digestive and Kidney Diseases, National Institutes of Health, Building 12A, 12 South Drive, Bethesda, MD 20892, USA
[*] Corresponding author

# Abstract

A nucleotide sequence 35 base pairs long can take 1,180,591,620,717,411,303,424 possible values. An example of systems biology datasets, protein binding microarrays, contain activity data from about 40000 such sequences. The discrepancy between the number of possible configurations and the available activities is enormous. Thus, albeit that systems biology datasets are large in absolute terms, they oftentimes require methods developed for rare events due to the combinatorial increase in the number of possible configurations of biological systems. A plethora of techniques for handling large datasets, such as Empirical Bayes, or rare events, such as importance sampling, have been developed in the literature, but these cannot always be simultaneously utilized. Here we introduce a principled approach to Empirical Bayes based on importance sampling, information theory, and theoretical physics in the general context of sequence phenotype model induction. We present the analytical calculations that underlie our approach. We demonstrate the computational efficiency of the approach on concrete examples, and demonstrate its efficacy by applying the theory to publicly available protein binding microarray transcription factor datasets and to data on synthetic cAMP-regulated enhancer sequences. As further demonstrations, we find transcription factor binding motifs, predict the activity of new sequences and extract the locations of transcription factor binding sites. In summary, we present a novel method that is efficient (requiring minimal computational time and reasonable amounts of memory), has high predictive power that is comparable with that of models with hundreds of parameters, and has a limited number of optimized parameters, proportional to the sequence length.

# Introduction

Experimental advances in systems biology produce an enormous amount of data relating nucleotide or protein sequences to quantitative measurements of biological activity. However, extracting useful/predictive information from this data remains a challenging problem. These datasets, often involving tens of thousands of sequences with associated phenotype measurements, are very large yet still sparse. For example, protein binding microarray technology (PBM) (Berger and Bulyk, 2006; Berger et al., 2006; Mukherjee et al., 2004; Philippakis et al., 2008) can produce more than 40000 35-mer DNA sequences. This is, however, negligible compared to the full sequence space of $4^{35}$ sequences. Thus, quantitative analysis of such datasets requires developing techniques that are effective for rare events and are computationally feasible, as well as limiting the complexity of models to avoid over-modeling the data.

The problem of over-fitting data can be approached from a Bayesian perspective. Empirical Bayes (Efron, 2010) as a term covers a large number of techniques. A general characteristic of these techniques is the use of the data itself to fix some parameters of the prior on the space of models, motivated by the properties of the James-Stein estimator (James and Stein, 1961). On the other hand, the question of a sparse sampling of sequence space can be addressed with importance sampling (Gelman and Meng, 1998; Hammersley and Morton, 1954; Woudt et al., 2007), a statistical technique used in many contexts to weight data points for computational efficiency and efficacy, especially in the presence of rare events.

The fundamental commonality among all inference techniques is information theory. We show that methods from statistical physics provide a simple and elegant way to combine information theory, importance sampling and Bayesian model comparison, enabling computationally efficient quantitative predictions of sequence-phenotype relationships.

We apply Bayesian model selection to the choice of data point weights, applying techniques used for computing effective actions or free energies in theoretical physics to link the importance sampling weighting of data points and the a priori likelihood of the weighting itself. Of course, using physical intuitions is not a novel approach for importance sampling (see (Gelman and Meng, 1998)), but the combination we consider here does not appear to have been considered before.

The nexus connecting information theory, theoretical physics, and importance sampling comes from the statistical physics of magnetic spins. Given a set of spin configurations written as vectors denoted $\vec{\sigma}$, an external inhomogeneous magnetic field vector, $\vec{h}$, can be used to align the spins so the magnetization vector, $\vec{m}$, the expectation value of the spin vector over all spin configurations, takes the specified (vector) value. This happens because $\vec{h}(\vec{m})$ is tuned such that spin configurations that lead to the desired magnetization $\vec{m}$ are given a higher or lower weight in the partition sum $Z[\vec{h}] = \sum_{\sigma} \exp(-\beta E(\vec{\sigma}) + \vec{h} \cdot \vec{\sigma})$ where $\beta$ is the inverse temperature and $E$ is the energy of $\vec{\sigma}$. The logarithm of this partition sum as a function of the magnetic field is $W[\vec{h}]$, the connected correlation generating function. By definition, then,

$$1 = \sum_{\sigma} \exp\left(-\beta E(\vec{\sigma}) + \vec{h} \cdot \vec{\sigma} - W[\vec{h}]\right). \tag{1}$$

In turn, $W[\vec{h}]$ is related to the Helmholtz free energy $F[\vec{m}]$ via the Legendre transform and the relation $\partial F / \partial \vec{m} = \vec{h}(\vec{m})$,

$$W[\vec{h}] + F[\vec{m}] = \vec{h} \cdot \vec{m}. \tag{2}$$

From eq. (2), we have $W[\vec{h}] = \vec{h} \cdot \vec{m} - F[\vec{m}]$, so replacing $W[\vec{h}]$ in eq. (1), we obtain

$$1 = \sum_\sigma e^{-\beta E} \exp\left(\vec{h} \cdot (\vec{\sigma} - \vec{m}) + F[\vec{m}]\right) \equiv \sum_\sigma e^{-\beta E} P(\vec{\sigma}),$$

defining the probability $P$ assigned to any configuration, depending only on the external magnetic field vector and the spin configuration vector. Notice that we distinguish between the weight (i.e., the importance sampling probability) of a configuration $P(\vec{\sigma})$ and the activity of the configuration $e^{-\beta E(\vec{\sigma})}$, as for example in noncentral hypergeometric distributions (Johnson et al., 2005). The Shannon entropy (Shannon, 1948) for a set of probabilities is defined as $-\sum_\sigma e^{-\beta E} P(\vec{\sigma}) \ln P(\vec{\sigma}) = -F[\vec{m}]$ and the expectation value of any quantity $x$ is defined as $\langle x \rangle = \sum_\sigma x e^{-\beta E} P(\vec{\sigma})$. We see then that the free energy as a function of the magnetization $\vec{m}$ is the negative of the Shannon entropy. Maximizing the Shannon entropy of the data as a function of $\vec{m}$ is thus the same as minimizing $F[\vec{m}]$, $\partial F / \partial \vec{m} = \vec{h}(\vec{m})$. There are no constraints, other than constraints implicit in the definition of the spin variables. For example, the frequencies of the bases at each position in a set of aligned DNA sequences have to add up to unity. This leads to the constraint that the mean value of $\vec{h}$ at each position can be taken to vanish. We assume nothing about the energy as a function of the spins, except that it is a polynomial function of the spin variables. The polynomial assumption is required because we will deduce a polynomial expansion for this energy, and a non-analytic energy function would probably be outside the validity of our method. It should be noted that every calculation in this paper has finite sums because all experimental data sets consist of finite number of sequences. There are no thermodynamic limits of any sort and there is no subtlety in convergence or boundary conditions.

Going back to the concept of importance sampling that motivated our introduction of $\vec{h}$ to weight data points, we note that not all values of $\vec{m}$ are equally likely given the data. As a simple example, the magnetization of spins that take only binary values cannot exceed 1, and if every configuration observed has the spin at a specific position taking the value 1, no magnetic field choice can change the magnetization of that spin to anything other than 1. Here, we consider choices of $\vec{m}$ as importance sampling models, and take the Shannon entropy as a function of $\vec{m}$ as the likelihood function. This is the background that leads us to the theory and practice that we present in this paper, the combination of the Empirical Bayes prior probability for possible values of the magnetization, the models, with a maximum entropy data likelihood function given by the Shannon entropy of the data constrained to that specific magnetization.

Instead of providing just an abstract account of our technique, we present here the theoretical underpinnings as applied to a particular problem of wide interest, the prediction of transcription factors' (TFs) affinities to/interaction with genes of interest and locating their corresponding binding sites. We emphasize that there are additional data such as chromatin immunoprecipitation assays available to improve such TF binding predictions, but we are concerned here with a very general method in the bioinformatics of sequence-phenotype inference, with TF binding used here only as a simple concrete example of our method.

Several high-throughput experimental methods have been developed to obtain quantitative data on TF-DNA interactions (Geertz and Maerkl, 2010). The most often used in vitro data are obtained from protein binding microarray experiments (PBMs) (Berger and Bulyk, 2006; Berger et al., 2006; Mukherjee et al., 2004; Philippakis et al., 2008), which utilize a universal array containing de Bruijn sequences of order $k$ providing an unbiased set of sequences containing all possible motifs of length $k$ or less (Philippakis et al., 2008). Other methods that aim to dissect a TF via systematic mutations include synthetic saturation mutagenesis (Patwardhan et al., 2009) and the more recent massively parallel reporter assay (MPRA) developed by (Melnikov et al., 2012) which uses quantitative sequence activity models (QSAM) (Jonsson et al., 1993) to model their data.

Various computational methods have been developed to obtain TF binding sites and/or predict the affinity of a TF to a sequence of nucleotides. Most methods rely on position weight matrices (PWMs) (Stormo, 2000; Stormo et al., 1982), also called position specific scoring matrices (Turatsinze et al., 2008), to represent TF specificity. PWMs assume each position contributes independently to the overall affinity of a TF to a sequence. This assumption was defended by arguing that in most cases PWMs are a good enough approximation as contributions to the binding energy from higher order interactions are negligible compared to the independent contributions (Zhao and Stormo, 2011). This is, however, not always the case (Bulyk et al., 2002; Maerkl and Quake, 2007; Tomovic and Oakeley, 2007; Zhao et al., 2012), and methods were thus needed to also take into consideration more complex interactions. Models have been suggested to improve upon the PWMs by also including information from dinucleotide interactions (Siddharthan, 2010; Zhao et al., 2012) and/or utilizing models with biophysical parameters (Djordjevic et al., 2003; Zhao et al., 2012), or by introducing defined weighted features instead of the mononucleotides of PWMs (Sharon et al., 2008). Other methods search for presence or absence of all possible oligonucleotides of length $k$ ($k$-mers) and assign them scores independent of their position within a sequence (Annala et al., 2011; Ghandi et al., 2014). A study analyzing the results of the DREAM5 challenge (Weirauch et al., 2013) showed that PWMs did indeed perform as well as more complex models at predicting PBM data probe intensities for most TFs but a lot worse for 10% of the tested TFs such that the overall winner (Annala et al., 2011) was a $k$-mer based method.

Generally, the models described above have a large number of parameters that increase exponentially with the number of required nucleotide interactions or complexity. For example, parameters required to capture single nucleotide effects correspond to weights assigned to the occurrences of each possible nucleotide at each sequence position. If $N$ is the number of parameters required to capture mononucleotide interactions and one needs to capture $n$-nucleotide interactions, the required number of parameters is proportional to $N^n$. This makes it difficult to optimize such models, as they become computationally very expensive and risk over-fitting the data. Furthermore, optimizing the parameters specifying higher nucleotide interactions in such a model in general alters the optimized values of lower order interaction parameters determined without such interactions. In particular, this change implies that there is no canonical embedding of the set of lower order models in the set of models including higher order interactions. On the other hand, $k$-mer based methods require including every possible sequence combination of length $k$, which can become intractable for larger values of $k$.

Our aim is to arrive at models that (1) have as direct a connection to available data as possible, (2) are computationally tractable, and (3) have superior predictive power. In most approaches aiming to find an energy function or activity measuring a quantitative phenotype from a sequence (the argument of the function), there are parameters specifying the function. Such parameters are always, obviously, determined using available training data. Usually the determination of parameters requires optimization in the form of minimization of a cost function, and it is well-known that such optimizations are difficult in a computational sense for complex energy function landscapes. There are two interesting theoretical aspects of our construction that make the parameter determination problem in our approach feasible. Firstly, our method divides parameters into two types: the expectation values defining the model space and the interaction parameters. In particular, we will show that the interaction parameters associated with higher order interactions are directly computed from the training data, requiring no optimization, by construction, once the expectation values are determined. There is no approximation involved in this statement, other than the uncontrolled approximation inherent in deducing a model from a finite amount of training data in any approach, the so-called Black Swan problem. Turning now to the determination of the expectation values, given a set of symbol sequences with associated quantitative measurements, such as sequence activity, we derive an exact, simple algebraic iteration that allows us to compute these expectation value parameters with minimal computational effort, much less than parameter optimization using standard cost function minimization algorithms. To capture the interaction parameters, in our construction, we have no choice: These interaction parameters are canonically (i.e., with no other choice possible) determined as the values of specific correlation functions in the data partition sum with configuration weights specified completely by the expectation values defining the model. To summarize, our models are completely specified by the data in two steps: First, an iteration determines the expectation values (equivalently, the probability of every data point), and then the canonical calculation of correlation functions from the data with these probabilities gives all higher order interaction parameters.

While we present specific examples in detail, the sequence-phenotype map is central to many quantitative questions in systems biology. Our setup is flexible and can be defined based on any set of sequence features of interest, whether it is position-dependent mono, di, or tri-nucleotides, position-independent $k$-mers, or any other problem-specific features. In this paper, we apply our approach to both PBM data posted on the DREAM5 challenge (Weirauch et al., 2013) website and published quantitative activity measurements (Melnikov et al., 2012) of randomly mutated sequences.

We will start by introducing the details of our approach in *Methods* showing several possible model setups. In *Results*, we demonstrate our approach by applying it to two different types of data, PBMs and massively parallel quantitative enhancer activity measurements. The versatility of our approach allows distinct model setups for these different datasets. We show that our approach has high predictive power that is comparable to that of the highest performing methods submitted to the DREAM5 challenge, as determined by the DREAM5 Challenge website (Weirauch et al., 2013), and the QSAM method used in (Melnikov et al., 2012). Our aim in showing these comparisons is mainly to demonstrate that the efficiency of our method and its limited number of parameters does not affect its predictive performance. Then, for completeness, we demonstrate the extraction of motifs or binding sites from these data, and we examine the effect of data size on model predictions.

## Methods

*Model Setup*

The starting point for our analysis is a set of symbol sequences, $s$, with associated quantitative measurements, $w(s)$. We start by embedding these sequences into a family of frequency models with each model specified by a set of frequencies $\vec{f}$. A specific nucleotide sequence is a model $\vec{f}$ with all frequencies 0 or 1. In other words, each sequence is translated into a binary vector. How this binary vector is defined depends on the available data. For example, to capture position-dependent nucleotide interactions, we define the binary vector $\vec{f} \equiv \{f_{iA}\}$ as

$$f_{iA} = \begin{cases} 1 & \text{if symbol } A \text{ is at position } i \\ 0 & \text{otherwise} \end{cases} \quad (3)$$

noting that a symbol can represent mono-, di-, tri-, or higher order nucleotides.

On the other hand, to capture $k$-mer information independent of their positions, we define $\vec{f} \equiv \{\vec{f}_j\}$ as

$$\vec{f}_j = \begin{cases} 01 & \text{if } k\text{-mer } j \text{ is found in the corresponding sequence} \\ 10 & \text{otherwise} \end{cases} \quad (4)$$

The definition in eq. (4) is equivalent to that in eq. (3) when we use coordinates on sequence space such that each sequence is replaced by another sequence composed of two possible symbols ($AB$) with the number of positions equal to number of $k$-mers taken into consideration. Thus, if $k$-mer $j$ is found in the original sequence then the new sequence takes the symbol $B$ at position $j$ and the symbol $A$, otherwise.

*Deriving the free energy on the space of possible sequences*

Our aim is to derive a function on the space of possible sequences such that the value of the function at any sequence is the latter's activity. We will reach our goal indirectly by first computing the maximum of a specific Bayesian posterior probability distribution and then use the outcome to calculate the desired activity function. A Bayesian calculation requires a model prior and a data-likelihood. For the first, we assume that the nucleotides/symbols that appear at a given position in the sequences are randomly chosen with no correlation between positions and that just the frequencies of occurrence determine the activity of a sequence. We need a family of models that reflects this set of null hypotheses.

A natural probability distribution for continuous parameters such as frequencies is the Dirichlet distribution. This distribution is usually interpreted as a distribution on vectors (or sets) of frequencies as the domain of the distribution is given by sets of positive numbers between 0 and 1 that sum up to 1. The Dirichlet distribution is the conjugate prior of the multinomial distribution and requires its set of parameters $\{q_i\}$ to be between 0 and 1, and $\sum_i q_i = 1$. Thus, for our models, we take the model prior for a set of frequencies $\{f_{iA}\}$ given a set of background Empirical Bayes frequencies, $\{q_{iA}\}$, to follow the Dirichlet distribution

$$\Pr(\{f_{iA}\} | \{q_{iA}\}, M) = \frac{\prod_A \Gamma(Mq_{iA})}{\Gamma\left(\sum_A Mq_{iA}\right)} \prod_A f_{iA}^{Mq_{iA}} ,$$

where $M$ is a smoothing parameter. Each choice of a set of frequencies $\vec{f}$ represents a magnetization model of a family of binary vectors (the spin configurations) associated with sequences. We use $\vec{f}$ instead of $\vec{m}$ because in the present context the numbers are really frequencies. Note that here the Dirichlet distribution is written with respect to $d\vec{f}/\vec{f}$.

We take the likelihood function, which is the probability of obtaining the set of observed sequences $\{s\}$ with expected frequencies $\vec{f} \equiv \{f_{iA}\}$, to be $\Pr(\{s\} | \vec{f}, M) = e^{-MF(\vec{f})}$, where $F(\vec{f})$ is the free energy evaluated at $\vec{f}$. To compute it, we introduce a source $\vec{h}$ and define the partition sum

$$Z \equiv \exp(MW(\vec{h})) = \sum_s \exp\left(-\beta E(s) + \sum_{i,A} Mh_{iA}\delta_{s_i,A}\right),$$

where $W(\vec{h})$ is the connected generating function, $\delta_{s_i,A}$ is 1 if position $i$ of sequence $s$ has nucleotide $A$ and is 0 otherwise, and $E(s)$ is the energy of $s$ defined as

$$e^{-\beta E(s)} = w(s)^\beta,$$

where $\beta = 1/T$ for temperature $T$. The expectation value of $E$ is then obtained via

$$M\frac{\partial W}{\partial \beta} = -\langle E \rangle.$$

Here, $\langle x \rangle \equiv \sum_s x\rho_s$ for weights of the probability distribution $\rho_s = e^{-\beta E(s)+M\vec{h}\cdot\vec{\sigma}_s} \Big/ \sum_s e^{-\beta E(s)+M\vec{h}\cdot\vec{\sigma}_s}$ and $\vec{\sigma}_s \equiv \{\delta_{s_i,A}\}$.

Then, $F(\vec{f})$ is the Legendre transform of $W(\vec{h})$ with respect to $\vec{h}$. In general, the Legendre-Fenchel transform is defined by $F(\vec{f}) = \sup_{\vec{h}}(\vec{h}\bullet\vec{f} - W(\vec{h}))$. A direct evaluation of this supremum is an optimization problem for the vector $\vec{h}$, for a given value of $\vec{f}$. Note that our method entails only this optimization. Once the vector $\vec{h}$ is determined as a function of $\vec{f}$, all higher orders in the effective action are directly determined by computing weighted one-particle irreducible correlation functions. However, in the latter part of our derivation we will show that even this optimization can be greatly simplified. When $W(\vec{h})$ is analytic in $\vec{h}$, we can use the simpler form:

$$F(\vec{f}) = \vec{h} \cdot \vec{f} - W(\vec{h}), \tag{5}$$

where $\vec{f} = \partial W/\partial \vec{h}$. Only this last relation between $\vec{f}$ and the derivative of $W$ is invalid if $W$ is not differentiable as a function of $\vec{h}$, which cannot be the case for finite datasets. On the other hand, the relation

$$M \frac{\partial F}{\partial \beta} = \langle E \rangle \tag{6}$$

is independent of this differentiability. This equation states that the derivative of the free energy with respect to $\beta$ evaluated at any specific model is given by the expectation value of $E$ in an ensemble where the expectation value of every nucleotide at each position is given by the frequencies defining that model. $\langle E \rangle$, computed using eq. (6), is a good predictor of $E(s)$ in the limit of small $\beta$, as we will show in Results. A small $\beta$ insures that the low activity sequences are given enough importance. On the other hand, increasing $\beta$ emphasizes higher activity sequences by giving them a higher weight compared to the bulk.

To calculate $\langle E \rangle$ from eq. (6) we still need to obtain the free energy at any frequency model $\vec{f}$. Now the Bayesian posterior distribution is obtained using Bayes' theorem:

$$\Pr(\vec{f} \mid \{s\}, M) = \frac{\Pr(\vec{f} \mid \{q_{iA}\}, M) \Pr(\{s\} \mid \vec{f}, M)}{\Pr(\{s\} \mid M)},$$

where the probability of the data, $\Pr(\{s\} \mid M)$, is a normalizing constant.

The most likely posterior model $\vec{f}^0$ maximizes $\Pr(\vec{f}|\{s\},M)$. If we have $\vec{f}^0$, we can obtain the free energy at other frequency models by a Taylor expansion around $\vec{f}^0$ with all derivatives determined directly from the data,

$$F(\vec{f}) = F(\vec{f}^0) + (\vec{f}-\vec{f}^0)\frac{\partial F}{\partial \vec{f}}\bigg|_{\vec{f}^0} + \frac{1}{2}(\vec{f}-\vec{f}^0)\frac{\partial^2 F}{\partial \vec{f}^2}\bigg|_{\vec{f}^0}(\vec{f}-\vec{f}^0)^T + \cdots \quad (7)$$

Define a function $K(\vec{f})$ such that $\Pr(\vec{f}|\{q_{iA}\},M) = e^{MK(\vec{f})}$. We take the probabilities $\{q_{iA}\}$ to be the frequencies found in the set of sequences $\{s\}$ independent of the activity of the sequences. Maximizing $\Pr(\vec{f}|\{s\},M)$ yields

$$\frac{\partial K}{\partial f_{iA}}\bigg|_{\vec{f}^0} = \frac{\partial F}{\partial f_{iA}}\bigg|_{\vec{f}^0}. \quad (8)$$

The left hand side of eq. (8) can be calculated directly from $e^{MK(\vec{f})}$, obtaining

$$\frac{\partial K}{\partial f_{iA}} = \frac{q_{iA}}{f_{iA}}.$$

By definition, eq. (5) requires $\frac{\partial F}{\partial f_{iA}}\bigg|_{\vec{f}} = h_{iA}$. Hence, at the most likely model, $\vec{f}^0$,

$$h_{iA}(\vec{f}^0) = \frac{\partial K}{\partial f_{iA}}\bigg|_{\vec{f}^0} = \frac{q_{iA}}{f_{iA}^0}.$$

With this re-organization, we can iteratively solve for $h_{iA}(\vec{f}^0)$,

$$\vec{f}_E^{(n+1)} = \vec{f}\left(\vec{h} = \frac{\partial K}{\partial \vec{f}}\bigg|_{\vec{f}_E^{(n)}}\right) \quad (9)$$

where $\vec{f}_E^{(n)}$ is the value of $\vec{f}$ at the $n^{th}$ iterative step. At the fixed point, we obtain $\vec{f}^0$ and in turn $\left.\frac{\partial F}{\partial f_{iA}}\right|_{\vec{f}^0}$ which is equal $h_{iA}(\vec{f}^0)$.

Turning back to the general case of the Legendre-Fenchel transform, we note that finding the model with the highest a posteriori probability is a matter of computing a supremum over $\vec{f}$ and over $\vec{h}$. Interchanging the two supremum evaluations, we are left with the Legendre-Fenchel transform of an explicitly known function, $K$, which is independent of the data and simple to compute a priori, and an optimization problem, but one involving the convex and analytic function $W$ and the Legendre-Fenchel transform of $K$, an explicitly computable function:

$$\begin{aligned}\sup_{\vec{f}}(K-F) &= \sup_{\vec{f}}\left(K - \sup_{\vec{h}}(\vec{f}\cdot\vec{h} - W)\right) \\ &= \sup_{\vec{f}}\sup_{\vec{h}}(K - \vec{f}\cdot\vec{h} + W) \\ &= \sup_{\vec{h}}(\hat{K} + W)\end{aligned}$$

where $\hat{K} = \sup_{\vec{f}}(K(\vec{f}) - \vec{f}\cdot\vec{h})$. Now $K(\vec{f})$ is explicitly differentiable and we can compute $\hat{K}(\vec{h})$ (numerically, if necessary) so that we are left with $\partial W/\partial \vec{h} = \langle\sigma\rangle_{\vec{h}}$ and $\partial \hat{K}/\partial \vec{h} = -\vec{f}(\vec{h})$ to solve for the supremum. We will not need this general case for the applications we present in this paper.

All derivatives of the free energy at $\vec{f}^0$ can now be directly computed, limited only by the number of sequences available. Here we derive the second order expansion as follows: since

$$\left.\frac{\partial^2 F}{\partial f_{iA}\partial f_{jB}}\right|_{\vec{f}^0} = \left.\frac{\partial h_{iA}(\vec{f})}{\partial f_{jB}}\right|_{\vec{f}^0} \text{ and } f_{jB} = \partial W/\partial h_{jB}, \text{ we obtain}$$

$$\frac{\partial^2 F}{\partial f_{iA}\partial f_{jB}} = \left[\frac{\partial^2 W}{\partial h_{iA}\partial h_{jB}}\right]^{-1},$$

and from $\partial^2 Z / \partial h_{iA} \partial h_{jB}$,

$$\frac{\partial^2 W}{\partial h_{iA} \partial h_{jB}} = M \left\{ \left\langle \delta_{s_i,A} \delta_{s_j,B} \right\rangle - \left\langle \delta_{s_i,A} \right\rangle \left\langle \delta_{s_j,B} \right\rangle \right\}.$$

The Taylor expansion is limited by the number of terms that we can reliably compute from the data, though each derivative is determined by the data at the fixed point due to the determination of the weights of the probability distribution. While the Taylor expansion is certainly easily computed, its numerical convergence to the exact form from the Legendre-Fenchel version is not guaranteed. We shall mainly use the first derivative and occasionally the second derivative in this paper. Note that a first-order Taylor approximation is a global lower bound for convex functions.

# Results

We demonstrate our method and assess its predictive power by applying it to model two types of data on transcription factor (TF) binding affinities.

## I- *Application to protein binding microarray (PBM) data*

To demonstrate our method we obtained data for 86 diverse mouse TFs from the datasets published in the DREAM5 challenge entitled "TF-DNA Motif Recognition Challenge" (Weirauch et al., 2013). These datasets consist of PBMs wherein sequences are assayed using two arrays of different 10-mer de Bruijn sequences, HK (Philippakis et al., 2008) and ME (Mintseris and Eisen, 2006). The quantitative measurements (i.e., $w(s)$) here are the mean probe intensities for each sequence $s$. The aim of the challenge was to use data from one array type (i.e., either HK or ME) to predict the other. In the challenge, the mean probe intensities are provided (along with other measurements that are not relevant to our method) for both array types in only 20 TFs, but only for one type in the rest of the 66 TFs. In what follows, we will first demonstrate how we set up our models and choose our parameters, and then use the method to predict the unknown intensities for the 86 TFs, showing that our method performs equally well (and even slightly better) than all the methods submitted for the competition.

### *Defining the set of models $\vec{f}$*

Here we set up two sets of models $\vec{f}^{\,k}$ and $\vec{f}^{\,p}$ where the first set captures $k$-mer information while the second captures position-dependent nucleotide interactions.

As the type of data we are dealing with ensures the presence of multiple copies of every non-palindromic $k$-mer of length 8 or less (Weirauch et al., 2013), it is intuitive to design our models using $k$-mer information. However, we do not want to use every single possible $k$-mer of each length $k \leq 8$ as these would lead to very large vectors. While we include all $k$-mers of length 6, for longer $k$-mers ($k \geq 7$) we only need to include those that give the most information on high intensity probes. For this, we calculate the Kullback-Leibler (KL) divergence between the frequency of each $k$-mer, $P_j = \sum_s \delta_{js} / \sum_j \sum_s \delta_{js}$ (where $\delta_{js}$ is 1 when $k$-mer $j$ is in sequence $s$ and 0 otherwise), and its intensity-weighted frequency, $\tilde{P}_j = \sum_s w(s)\delta_{js} / \sum_j \sum_s w(s)\delta_{js}$:

$$KL(P_j, \tilde{P}_j) = P_j \ln\left(\frac{P_j}{\tilde{P}_j}\right) + (1-P_j)\ln\left(\frac{1-P_j}{1-\tilde{P}_j}\right). \tag{10}$$

The most informative $k$-mers are those with the highest KL values. We found that there is no difference between our predictions whether we include only the highest 200 or the highest 300 7-mers and 8-mers (see supplementary figure S1 for typical examples). On the other hand, if we include too many $k$-mers we run the risk of overfitting the data. When data are overfit, we observe a large improvement in fitting the training set while the fit over the test set becomes worse. We set up the set of models $\vec{f}^k$ using the definition in eq. (4).

For the second set of models, $\vec{f}^p$, we use the definition in eq. (3), where each symbol represents a tri-nucleotide. It is possible for these datasets to include tri-nucleotide interactions as each possible 3-mer is represented more than once at each position.

*Choice of $M$ and $\beta$*

The smoothing parameter $M$ can be thought of as a measure of confidence in the data, as the Dirichlet distribution becomes singular in the limit $M = 0$ which corresponds to 100% confidence. Here, we use approximately the highest value of $M$ that leads to a solution for the optimal model $\vec{f}^0$, as eq. (9) does not numerically converge when $M$ is too high. For the results shown here we use $M \simeq 0.1$. The final results, however, are not sensitive to small changes in $M$ (see supplementary figure S2 for typical examples).

As discussed in **Methods**, the parameter $\beta$ needs to be small enough for $\langle E \rangle$ to be a good prediction of $E$. However, if $\beta$ is too small, it gives too little weight to high intensity sequences. To insure we correctly predict the few high intensity sequences without ignoring the bulk of low and medium intensities, we first run our algorithm at $\beta = 0.1$ using the set of models $\vec{f}^k$, obtaining our first predictions $\langle E_1 \rangle$. On the residuals, $E - \langle E_1 \rangle$, we run the algorithm again at $\beta = 0.01$, obtaining $\langle E_2 \rangle$. On the residuals of this second run, $E - \langle E_1 \rangle - \langle E_2 \rangle$, we run the algorithm once more using the set of models $\vec{f}^p$ at $\beta = 0.01$. $\langle E \rangle$ at each run is obtained via eq. (6) where $F$ is approximated by the Taylor expansion in eq. (7) up to the first term,

$$F(\vec{f}) \approx F(\vec{f}^0) + (\vec{f} - \vec{f}^0) \frac{\partial F}{\partial \vec{f}}\bigg|_{\vec{f}^0}.$$

*Rectifying the histogram of the final prediction*

As our aim is to demonstrate our algorithm, we looked for some benchmarks to validate the efficacy of our procedure. To obtain a fair comparison, we had to account for the vicissitudes of the data, as did all the participants in the original challenge. We benefited from a close study of the work of the winners of this challenge, who showed that these specific datasets have many probes that were saturated at high intensities and corrected for this artifact by preprocessing the data (Annala et al., 2011). We did not perform any preprocessing as we have no expertise in the sophisticated techniques used for this (see (Annala et al., 2011)). Undoubtedly, if we did, our results would improve, as demonstrated by the winning entry (Annala et al., 2011), but here, instead, we simply rectify the histogram of our final predictions. The idea is that saturation will cause a few predictions to be much lower or much higher than they should be. These few outliers will artificially skew the values of the Pearson correlation between the predicted intensities and the actual ones. Here, we rectify the histogram of our final predictions by moving isolated points in the highest/lowest bins into lower/higher empty bins (Figure 1 a and b). This minor manipulation keeps the ranking intact and does not change the values of the Pearson correlation between the predicted log intensities and that of the actual ones (Figure 1 c).

*Algorithm performance at predicting PBM intensities*

To assess the predictive power of our method on PBM data, we first applied it to the 20 TFs where both HK and ME data were given. We trained our algorithm using the mean intensities of sequences from one array type (HK or ME) and predicted those of sequences from the other type (ME or HK). In table 1 we show the values of the Pearson correlations between HK and ME log intensities (Pearson log) and between intensities (Pearson and Spearman) correlations. For all TFs, we used $M = 0.1$ except for Junb where we used $M = 0.08$ as eq. (9) did not converge at $M = 0.1$ for this TF. Decreasing the value of $M$ also decreases the CPU time, thus longer convergence times indicate that the value of $M$ is reaching the non-convergence limit. We compare our results with that of team D, the DREAM5 challenge winners (table 1 columns 5 and 7). Here we only compare our Pearson and Spearman values as these are the scoring values provided in (Annala et al., 2011). We obtained the same average Spearman value and a higher Pearson value with an average CPU time of 12 minutes per TF (~44000 sequences in the training set).

We next ran our algorithm on the remaining 66 TFs and uploaded our predicted intensities to the DREAM5 challenge website (Weirauch et al., 2013). The scoring was calculated based on five evaluation criteria: Pearson of the log values, Pearson, Spearman, area under the receiver operating characteristic (AUROC) of the 8-mer values, and the area under the precision-recall curve (AUPR) of the 8-mer values. Again, our method ranked well compared to challenge participants (table 2). Of course, this comparison is not to suggest in any way that we would have fared as well without the benefit of knowing the results and methods of the participants in the challenge. Similarly, another method, FeatureReduce, which was not represented among the algorithms submitted to the challenge, was later shown to perform slightly better than the winner's algorithm (Weirauch et al., 2013). Unfortunately, this method was not included on the website and we were thus not able to include it in our comparison. Regardless, our main aim in presenting the comparison here is to provide an unbiased evaluation of the efficacy of our approach.

*Motif discovery*

TF binding motif discovery is a problem of great interest (Zambelli et al., 2013) and was one of the aims of the DREAM5 challenge (Annala et al., 2011; Weirauch et al., 2013). Here, for completeness, we show how one can obtain sequence motifs from the PBM data using our method. Above, we showed that, for this particular data, 200 8-mers chosen based on their KL divergence values, eq. (10), were enough to provide the information needed to predict very high (and in some cases also very low) intensity sequences. Therefore, this set should suffice to obtain the TF binding motif. We first ran our algorithm once using the $\vec{f}^k$ set of models where only the 200 8-mers with the highest KL divergence value are included (this takes only a few seconds of CPU time). With this, we obtain $\vec{h} \equiv \{h_j\}$ at the fixed point for $j = 1...200$. The 8-mers most responsible for the high intensity of a sequence are those with the most negative $h$ value while those responsible for lowering the sequence intensity have a positive value of $h$. Since we are only interested in the former, for motif finding purposes, we removed all 8-mers with a positive $h$ value. We aligned the rest using Clustal Omega (Goujon et al., 2010; Sievers et al., 2011). We then prepared 20000 sequences where the 8-mers are repeated several times with probability $e^{-h}$ and obtained a logo representing the motif using WebLogo 3 (Crooks et al., 2004).

To demonstrate the effectiveness of this procedure we tested it on 11 out of the 20 known TFs and compared our logos with those obtained from the JASPAR database (Figure 2). These 11 TFs were the only ones that we were able to find in the database. In Figure 2, we only show 8 representative ones to insure visibility. We found that though our logos are generally longer than the reference ones in JASPAR, they all (with the exception of a few) contained sections with almost identical motifs to the reference motifs. The quality of the motif does not improve if we also include the most informative 200 7-mers (supplementary figure S3) and becomes worse when we include 200 7-mers and 1000 6-mers (supplementary figure S4). This shows that for this data, high intensity sequences are mostly dictated by the 8-mers present.

*Finite size effect*

Here we assess the effect of the sample size on the goodness of fit. We chose two representative TFs, one with a high Pearson Log value, Foxo4, and another with a low Pearson Log value, Zscan20. In each case we used the full ME data as the test set and evaluated the goodness of fits after training our models with a randomly chosen part of the HK data. For each set number of sequences we performed 10 different trials and reported the mean values of the Pearson Log, Spearman, AUROC, and PR for the 35-mer probes (Figures 3 and 4). In each case the goodness of fit increases with increasing number of sequences in the training set. However, the difference is not big. Pearson Log value increased by only $0.0260\pm0.0020$ for Foxo4 and $0.0150\pm0.0027$ for Zscan20 when using only 40% of sequences in the HK data vs using the full set (Figures 3a and 4a). Similarly, the differences between the Spearman values, AUROC, and PR for the 35-mer probes were $0.0217\pm0.0019$ (Figure 3b), $0.0031\pm0.0013$ (Figure 3c),

$0.0294 \pm 0.0131$ (Figure 3d), respectively for Foxo4, and $0.0122 \pm 0.0025$ (Figure 4b), $0.0157 \pm 0.0027$ (Figure 4c), $0.1484 \pm 0.0792$ (Figure 4d), respectively for Zscan20.

We next ask if increasing the number of sequences can lead to much better results. For this, we fit the inverse of the mean values of the Pearson Log and Spearman with $aN^{-\alpha} + b$ where $a$, $b$, and $\alpha$ are constants to be determined and $N$ is the number of sequences used for training. For Foxo4, we obtain $\frac{1}{Pearson\ Log} = 75.259 N^{-0.7} + 1.1672$ ($r^2 = 0.9989$) which implies that the value of the Pearson Log goes to 0.857 as $N \to \infty$, and $\frac{1}{Spearman} = 3.139 N^{-0.3} + 1.2026$ ($r^2 = 0.9969$) which implies that the Spearman value goes to 0.832 as $N \to \infty$. For Zscan20, we obtain $\frac{1}{Pearson\ Log} = 96855 N^{-1.3} + 3.7847$ ($r^2 = 0.9923$) which implies that the value of the Pearson Log goes to 0.264 as $N \to \infty$, and $\frac{1}{Spearman} = 628688 N^{-1.5} + 4.2911$ ($r^2 = 0.9898$) which implies that the Spearman value goes to 0.233 as $N \to \infty$. Of course, these asymptotic values we obtained are based on phenomenological models with the existence of an asymptotic value assumed by definition. Most importantly, these values are only meaningful under the assumptions that an increase in the number of training set sequences only provides more of the same type of information and that the model setups stay the same. For example, a much larger number of sequences might provide enough information on longer $k$-mers, making it possible to include them in the setup. Alternatively, a larger training set might make it more feasible to consider higher-order $k$-mer interactions which would undoubtedly improve the goodness of fit, particularly in cases where there are more complex interactions between different binding sites.

*II-    Application to massively parallel reporter assays (MPRA)*

To ensure that the performance of our method is not restricted to the type of data presented above, we also applied it to randomly mutated sequences of a synthetic cAMP-regulated enhancer (CRE), and of the human interferon-β1 enhancer (Ifnb), with quantitative activity measurements obtained from MPRA generated by (Melnikov et al., 2012).

### *Defining the set of models $\vec{f}$ for CRE*

We randomly divided the data into training set and test set. As above, we used the former to obtain $\langle E \rangle$ and the latter to gauge the accuracy of our predictions of the activity of new sequences from $\langle E \rangle$. The number of sequences available in this data set is 26337. After removing 10%, 25%, or 50% of data for testing we are left with only 23704, 19753, or 13169 sequences in the training set. This necessitates decreasing the number of vector components we use to describe the sequences in order to avoid over-fitting.

As with the PBM data above, we set up two sets of models $\vec{f}^k$ and $\vec{f}^p$ to capture both position-independent $k$-mer information and position-dependent nucleotide interactions, respectively. Here, we only used the 400 6-mers with the highest KL values to set up $\vec{f}^k$. For the set of models $\vec{f}^p$, we use the definition in eq. (3) where each symbol represents a single nucleotide. The 3-mer representation is inappropriate here both because there is not enough data and because the sequences are mutations from a reference, making it less likely to cover all possible 3-mers at each position. For all trials (results shown in tables 3, 4, 5) we used $M = 0.1$ and ran the algorithm two times (as for the PBM data) using the set of models $\vec{f}^k$ with $\beta = 0.2$ for the first run, and $\beta = 0.1$ for the second, then a third time using the set of models $\vec{f}^p$ with $\beta = 0.01$. Note that these results are not sensitive to small changes in the choice of $\beta$ values (see supplementary table S1), nor to the $M$ value, chosen so that the algorithm converges at each step, as explained in the *Choice of $M$ and $\beta$* section.

$F$ is approximated by the Taylor expansion of eq. (7) up to the first term for the first two runs and up to the second term, $F(\vec{f}) \approx F(\vec{f}^0) + (\vec{f} - \vec{f}^0) \frac{\partial F}{\partial \vec{f}}\bigg|_{\vec{f}^0} + \frac{1}{2}(\vec{f} - \vec{f}^0) \frac{\partial^2 F}{\partial \vec{f}^2}\bigg|_{\vec{f}^0} (\vec{f} - \vec{f}^0)^{\mathrm{T}}$, for the last run with the position-dependent models.

*Algorithm performance at predicting CRE activity measurements*

Testing the predictive power of our method using 30 different trials (where data are randomly divided into a training set and a test set), we find that our model performs better than the linear quantitative sequence activity models (QSAM) (Melnikov et al., 2012) ($r^2 = 0.679 \pm 0.007$ for predicting the test sets compared with $r^2 = 0.63$ obtained using linear QSAM). The goodness of fit does not significantly change whether we use only 10% of the sequences for testing (table 3, $r^2 = 0.682 \pm 0.008$, Spearman = $0.844 \pm 0.006$), 25% (table 4, $r^2 = 0.682 \pm 0.006$, Spearman = $0.843 \pm 0.004$), or 50% (table 5, $r^2 = 0.674 \pm 0.004$, Spearman = $0.839 \pm 0.002$). Note that using a replicate of CRE (generated by (Melnikov et al., 2012)) as a training set to fit the above CRE data set, we obtain $r^2 = 0.726$ while with the opposite (i.e., using the above data as training to predict the replicate) we obtain $r^2 = 0.617$. We only compare our results with the linear QSAM obtained by (Melnikov et al., 2012) even though they did use multiple types of models to fit their data. For example, the highest $r^2$ they reported was $r^2 = 0.723$ using a linear/nonlinear model, but this was for training data only. However, they only reported a fivefold cross-validation for the linear QSAM and not for the other models, though the other models required more parameters.

*Defining the set of models $\vec{f}$ for Ifnb*

There are 26833 sequences available for the Ifnb data set. We randomly set aside 25% of the sequences for testing and use the others for training. Despite the fact that both CRE and Ifnb have approximately the same number of sequences, we found that the set of models used for CRE overfit the Ifnb data (as there was a big difference between the Pearson correlations obtained by fitting the training set and that obtained by fitting the test set). Instead, we need only use one run with one set of models $\vec{f}^p$ that captures position-dependent mononucleotides. We approximate $F$ by the Taylor expansion of eq. (7) up to the second term, and use $M = 0.1$ and $\beta = 0.01$.

This difference between the models used to fit CRE and those for Ifnb stems from the nature of the Ifnb data set. Many of these sequences show no appreciable activity change despite an increasing number of mutations and only a few sequences show enhanced activation or deactivation. On the other hand, the CRE sequences have a wide range of activities that are dependent on the sequences (see supplementary figure S5). This implies that most of the sequences in the Ifnb data set are uninformative and confound modeling.

*Algorithm performance at predicting Ifnb activity measurements*

Testing with 10 different trials (using 25% of sequences for testing), we obtained $r^2 = 0.047 \pm 0.006$ for predicting the test sets (table 6). Ifnb also showed much more variation between the different trials than CRE did, particularly in the values of their Pearson correlations (column 4 of table 6). For the training set on the full data we obtained $r^2 = 0.068$ using first term approximation for $F$ in the Taylor series, and $r^2 = 0.286$ using the second term approximation. (Melnikov et al., 2012) obtained $r^2 = 0.071$ using a linear QSAM and $r^2 = 0.104$ using a dinucleotide model. In both cases, the $r^2$ reported is that of the training set and no fivefold cross-validation was provided.

*Motif and binding sites discovery for CRE*

We first attempt to find a motif using the same procedure used for the DREAM5 TF data. We ran the algorithm once using the $\vec{f}^k$ set of models where only the 400 6-mers with the highest KL divergence values are included. We aligned the 6-mers with a negative $h$ and obtained the corresponding logo (Figure 5). The motif obtained corresponds to positions 65 to 83 which include a cryptic site and the fourth binding site, CREB(4), of the known CRE sequence (Figure 6). This is consistent with (Melnikov et al., 2012) results showing that CREB(4) has the largest contribution in enhancing the activity of the induced CRE. However this type of data is not meant for motif finding. There is no guarantee that a motif found following the above procedure will lead to interpretable results and sometimes only part of the obtained motif can be mapped to the reference sequence (see supplementary figure S6). On the other hand, this data set, measuring the activity of different mutations from a known reference sequence, allows one to see interactions between the different positions and thus extract the locations of the binding sites.

To ascertain the interactions that underlie the activity of sequences, we compute a mutual information matrix, $\{I_{ij}\}$, where

$$I_{ij} = \frac{1}{M} \sum_{A,B} \frac{\Pr(iA, jB)}{\Pr(*, *)} \ln\left(\frac{\Pr(iA, jB)\Pr(*, *)}{\Pr(iA, *)\Pr(*, jB)}\right)$$

is the mutual information between positions $i$ and $j$ where $i, j = 0, 1, \cdots N_p$ for $N_p$ positions, and $\Pr(iA, jB)$ is the joint probability distribution of a perturbation $\vec{\varepsilon}^{(iA, jB)}$ in the direction increasing $f_{iA}$ and $f_{jB}$ while keeping the sum of frequencies equal to one at each position:

$$\Pr(iA, jB) \propto \exp\left\{-MF\left(\vec{f}^0 + \vec{\varepsilon}^{(iA,jB)}\right)\right\}, \Pr(iA, *) = \sum_B \Pr(iA, jB), \Pr(*, jB) = \sum_A \Pr(iA, jB), \text{ and}$$

$$\Pr(*, *) = \sum_{A,B} \Pr(iA, jB).$$

To obtain the binding sites, only high activity sequences need to be considered. We thus compute the mutual information matrix using only the 10% highest activity sequences with $M = 0.1$ and $\beta = 1.0$. The resultant matrix (Figure 7a) clearly shows four interaction sites that are consistent with the known binding sites of CRE (CREB(1), CREB(2), CREB(3), and CREB(4) in Figure 6). Note that, for small $\beta$, nucleotide interactions are distributed throughout the promoter sequence and do not highlight any localized sites. Similarly, increasing the number of lower activity sequences (i.e., decreasing the percent of high activity sequences contributing to the partition sum) results in the loss of the power of statistical averaging. For example, using only the highest 18% activity sequences, only three of the known binding sites appear in the resultant mutual information matrix (Figure 7b), two with 25% (Figure 7c), and only one binding site, CREB(4), with 50% (Figure 7d).

## Conclusions

The statistical physics of modeling (Brown and Sethna, 2003; Gutenkunst et al., 2007; Kinney et al., 2007; Weirauch et al., 2013) is an important aspect of extracting information from data in biology as the data are never complete, and the structure of possible models is never certain. Such systems are typically complex enough that they cannot be modeled with explicit parameterized models incorporating every interactant. Simplified effective model development requires trade-offs between goodness of fit and model complexity (Danuser et al., 2013; Samaga and Klamt, 2013). However, optimizing a model incorporating the mostly unknown interactions of even 1,000 genes, let alone 30,000, is practically impossible.

We developed a novel method to extract model information from sequence-phenotype data, introducing a number of parameters commensurate with the length of the sequence. Our method relies on three conceptual foundations: (1) the connection between the physical free energy and Shannon entropy; (2) the weighting of data as used in importance sampling or robust regression; and (3) computational efficiency enabled by model parameterization in terms of the expectation values, which, in particular, enabled the use of an iterative method to find the most likely posterior model. Even though the models are explicitly parameterized by this small number of parameters relative to the exponentially large numbers of parameters that one might naively introduce to take interactions between nucleotides at distinct sequence positions into account, determining these few parameters canonically dictates all such interaction terms directly from the data without any optimization. This last point is a major distinction between explicitly parameterized and optimized models and our approach.

The models we derive have a flexible initial setup such that the method can be applied to a variety of data, not only data on TF interactions but also any data wherein sequences are associated with a quantitative measurement of the phenotype of interest. In particular, the models are described in the same variables as the sequence data, avoiding the disconnect between experimental observations and theoretical model variables mentioned above. As befits an approach that is motivated by importance sampling, we showed that our results approached asymptotic values with relatively few sequences. The computational efficiency of our approach is demonstrated by the CPU times given in Tables 1, 3-6, for example. The values we report in Table 1 can be compared to the methods used in the DREAM5 challenge where the top 3 algorithms took less than 24 hours of CPU time while some others took over a week of CPU time (Weirauch et al., 2013).

A feature and a limitation of our analysis is that the free energy computed is convex. On the positive side, a linear approximation, as used in this paper, is a global lower bound for convex functions. On the other hand, the iteration underlying the efficient computation of the maximum a posteriori model fails to converge if convexity is violated. The complete Legendre-Fenchel transform results in a more standard parameter optimization problem, albeit with the smaller number of parameters involved in our approach. However, our accuracy in predicting activity will be limited by this obligatory convexity. In particular, when the underlying data consists of two separate clusters, we have found that our method can be used to separate the clusters (data not shown), and then must be applied again to each cluster independently. The importance sampling of sequences enforced by our iteration is determined by our choice of model distribution. We have verified that our results do not depend on varying this distribution.

Our method is widely applicable to any data that can be put into the format of sequence-phenotype pairs. We showed explicitly that our approach has better or equal predictive power compared to both simple models with a number of parameters linear in the sequence length and to those accounting for interactions between distal sequence positions. If the phenotype is discrete, a variety of techniques have been developed in the statistical literature to regress against categorical data, so the continuity of the phenotype in our transcription factor examples is not a prerequisite for using our method. It is computationally efficient, requiring seconds to minutes on a desktop workstation to extract the Shannon entropy from tens of thousands of sequence-activity pairs. The dependence of this Shannon entropy on the frequency expectation value pinpoints the features (e.g., positions in a sequence of nucleotides) that interact to affect a particular quantitative phenotype.

## Acknowledgments

This work was supported by Intramural Research Program of the National Institutes of Health, NIDDK. We thank Arthur Sherman, Carson Chow and Terry Hwa for helpful conversations.

**Table 1**

| TF | array type | Pearson Log | Pearson | Pearson (teamD) | Spearman | Spearman (team D) | CPU time (mins:secs) |
|---|---|---|---|---|---|---|---|
| Cebpb | HK --> ME | 0.712 | 0.658 | 0.524 | 0.674 | 0.657 | 9:42 |
|  | ME --> HK | 0.767 | 0.664 | 0.494 | 0.755 | 0.773 | 11:00 |
| Egr2 | HK --> ME | 0.783 | 0.768 | 0.798 | 0.734 | 0.735 | 14:51 |
|  | ME --> HK | 0.728 | 0.712 | 0.726 | 0.712 | 0.711 | 15:55 |
| Esr1 | HK --> ME | 0.650 | 0.676 | 0.746 | 0.575 | 0.679 | 9:30 |
|  | ME --> HK | 0.447 | 0.537 | 0.607 | 0.374 | 0.384 | 12:27 |
| Foxj2 | HK --> ME | 0.689 | 0.556 | 0.620 | 0.678 | 0.651 | 10:30 |
|  | ME --> HK | 0.720 | 0.570 | 0.590 | 0.709 | 0.706 | 13:33 |
| Foxo1 | HK --> ME | 0.576 | 0.620 | 0.619 | 0.473 | 0.482 | 10:15 |
|  | ME --> HK | 0.748 | 0.697 | 0.615 | 0.660 | 0.722 | 12:50 |
| Foxo3 | HK --> ME | 0.673 | 0.675 | 0.574 | 0.627 | 0.589 | 19:36 |
|  | ME --> HK | 0.713 | 0.679 | 0.727 | 0.706 | 0.740 | 10:44 |
| Foxo4 | HK --> ME | 0.825 | 0.743 | 0.731 | 0.749 | 0.713 | 9:27 |
|  | ME --> HK | 0.852 | 0.721 | 0.773 | 0.817 | 0.823 | 12:30 |
| Foxp1 | HK --> ME | 0.694 | 0.711 | 0.689 | 0.623 | 0.633 | 9:31 |
|  | ME --> HK | 0.765 | 0.685 | 0.550 | 0.729 | 0.724 | 13:47 |
| Foxp2 | HK --> ME | 0.740 | 0.599 | 0.693 | 0.738 | 0.741 | 13:03 |
|  | ME --> HK | 0.659 | 0.441 | 0.385 | 0.647 | 0.612 | 18:38 |
| Gmeb2 | HK --> ME | 0.917 | 0.213 | 0.808 | 0.880 | 0.857 | 10:06 |
|  | ME --> HK | 0.948 | 0.490 | 0.767 | 0.918 | 0.899 | 11:21 |
| Irf2 | HK --> ME | 0.794 | 0.550 | 0.725 | 0.741 | 0.746 | 16:00 |
|  | ME --> HK | 0.798 | 0.597 | 0.638 | 0.731 | 0.739 | 14:54 |
| Junb | HK --> ME | 0.585 | 0.578 | 0.479 | 0.593 | 0.570 | 8:51 |
|  | ME --> HK | 0.642 | 0.620 | 0.660 | 0.634 | 0.679 | 10:09 |
| Mecp2 | HK --> ME | 0.742 | 0.734 | 0.717 | 0.734 | 0.730 | 8:24 |
|  | ME --> HK | 0.810 | 0.791 | 0.801 | 0.804 | 0.817 | 15:06 |
| Nr2C1 | HK --> ME | 0.590 | 0.650 | 0.582 | 0.543 | 0.477 | 9:21 |
|  | ME --> HK | 0.549 | 0.556 | 0.495 | 0.491 | 0.433 | 10:12 |
| Pou3f1 | HK --> ME | 0.718 | 0.743 | 0.651 | 0.688 | 0.670 | 12:59 |
|  | ME --> HK | 0.746 | 0.677 | 0.463 | 0.683 | 0.680 | 14:09 |
| Sox14 | HK --> ME | 0.709 | 0.635 | 0.630 | 0.716 | 0.709 | 12:46 |
|  | ME --> HK | 0.731 | 0.671 | 0.695 | 0.716 | 0.726 | 14:03 |
| Sp1 | HK --> ME | 0.633 | 0.630 | 0.633 | 0.623 | 0.617 | 9:45 |
|  | ME --> HK | 0.640 | 0.637 | 0.641 | 0.622 | 0.634 | 9:51 |
| Tbx3 | HK --> ME | 0.669 | 0.661 | 0.438 | 0.626 | 0.645 | 8:54 |
|  | ME --> HK | 0.561 | 0.597 | 0.478 | 0.505 | 0.451 | 10:15 |
| Tcf3 | HK --> ME | 0.756 | 0.818 | 0.636 | 0.506 | 0.508 | 8:09 |
|  | ME --> HK | 0.737 | 0.663 | 0.467 | 0.488 | 0.525 | 10:42 |
| Zscan20 | HK --> ME | 0.262 | 0.270 | 0.239 | 0.233 | 0.225 | 18:59 |
|  | ME --> HK | 0.394 | 0.390 | 0.388 | 0.376 | 0.418 | 11:12 |
| **Average** |  | **0.692** | **0.622** | **0.612** | **0.646** | **0.646** | **12:04** |

**Table2**

| Team | Model type | Final Rank | Pearson | PearsonLOG | Spearman | AUROC | AUPR |
|---|---|---|---|---|---|---|---|
| our team | our method | 1 (1.8) | 0.643 (1) | 0.680 (2) | 0.640 (4) | 0.996 (1) | 0.713 (1) |
| Team_D | k-mer | 2 (2.8) | 0.641 (2) | 0.674 (3) | 0.639 (5) | 0.994 (2) | 0.7 (2) |
| Team_F | Other | 3 (4.6) | 0.61 (5) | 0.673 (4) | 0.655 (3) | 0.976 (5) | 0.545 (6) |
| Team_E | PWM | 4 (4.6) | 0.637 (3) | 0.694 (1) | 0.673 (2) | 0.952 (8) | 0.522 (9) |
| Team_G | k-mer | 5 (5.6) | 0.573 (7) | 0.621 (7) | 0.574 (7) | 0.994 (3) | 0.674 (4) |
| Team_J | Other | 6 (6.0) | 0.612 (4) | 0.65 (5) | 0.623 (6) | 0.965 (7) | 0.524 (8) |
| Team_I | Other | 7 (6.8) | 0.581 (6) | 0.647 (6) | 0.692 (1) | 0.94 (10) | 0.306 (11) |
| Team_H | Other | 8 (8.8) | 0.469 (11) | 0.417 (13) | 0.367 (13) | 0.991 (4) | 0.676 (3) |
| Team_C | Other | 9 (8.8) | 0.518 (9) | 0.523 (11) | 0.484 (11) | 0.975 (6) | 0.53 (7) |
| Team_9 | Other | 10 (9.4) | 0.497 (10) | 0.575 (8) | 0.562 (8) | 0.941 (9) | 0.248 (12) |
| Team_A | k-mer | 11 (10.0) | 0.533 (8) | 0.461 (12) | 0.431 (12) | 0.925 (13) | 0.584 (5) |
| Team_12 | k-mer | 12 (11.2) | 0.461 (12) | 0.544 (9) | 0.538 (9) | 0.929 (12) | 0.15 (14) |
| Team_K | k-mer | 13 (11.4) | 0.461 (13) | 0.54 (10) | 0.531 (10) | 0.93 (11) | 0.156 (13) |
| Team_B | PWM | 14 (13.2) | 0.267 (14) | 0.189 (14) | 0.1 (14) | 0.891 (14) | 0.462 (10) |
| Team_14 | PWM | 15 (15.0) | 0.0 (15) | 0.0 (15) | 0.0 (15) | 0.487 (15) | 0.003 (15) |

**Table3**

| trial | Pearson Log (r) | $r^2$ | Pearson | Spearman | CPU time (seconds) |
|---|---|---|---|---|---|
| 1 | 0.826 | 0.682 | 0.840 | 0.849 | 77 |
| 2 | 0.836 | 0.699 | 0.843 | 0.852 | 77 |
| 3 | 0.826 | 0.682 | 0.816 | 0.849 | 77 |
| 4 | 0.825 | 0.680 | 0.832 | 0.841 | 76 |
| 5 | 0.827 | 0.684 | 0.842 | 0.848 | 76 |
| 6 | 0.821 | 0.674 | 0.824 | 0.836 | 78 |
| 7 | 0.820 | 0.673 | 0.825 | 0.839 | 77 |
| 8 | 0.826 | 0.681 | 0.833 | 0.841 | 77 |
| 9 | 0.819 | 0.671 | 0.826 | 0.835 | 76 |
| 10 | 0.830 | 0.689 | 0.830 | 0.848 | 77 |
| Average | 0.826 | 0.682 | 0.831 | 0.844 | 77 |
| Standard Deviation | 0.005 | 0.008 | 0.009 | 0.006 | 0.6 |

**Table4**

| trial | Pearson Log (r) | r² | Pearson | Spearman | CPU time (seconds) |
|---|---|---|---|---|---|
| 1 | 0.829 | 0.688 | 0.830 | 0.847 | 65 |
| 2 | 0.821 | 0.675 | 0.826 | 0.838 | 66 |
| 3 | 0.828 | 0.685 | 0.834 | 0.842 | 66 |
| 4 | 0.832 | 0.692 | 0.837 | 0.849 | 67 |
| 5 | 0.823 | 0.677 | 0.823 | 0.840 | 68 |
| 6 | 0.827 | 0.685 | 0.827 | 0.846 | 67 |
| 7 | 0.821 | 0.673 | 0.824 | 0.836 | 67 |
| 8 | 0.826 | 0.682 | 0.830 | 0.844 | 67 |
| 9 | 0.823 | 0.678 | 0.829 | 0.841 | 67 |
| 10 | 0.826 | 0.683 | 0.828 | 0.843 | 67 |
| Average | 0.826 | 0.682 | 0.829 | 0.843 | 67 |
| Standard Deviation | 0.004 | 0.006 | 0.004 | 0.004 | 0.8 |

**Table5**

| trial | Pearson Log (r) | r² | Pearson | Spearman | CPU time (seconds) |
|---|---|---|---|---|---|
| 1 | 0.823 | 0.677 | 0.828 | 0.839 | 55 |
| 2 | 0.821 | 0.674 | 0.824 | 0.840 | 56 |
| 3 | 0.824 | 0.679 | 0.826 | 0.841 | 56 |
| 4 | 0.819 | 0.671 | 0.828 | 0.836 | 57 |
| 5 | 0.823 | 0.678 | 0.828 | 0.840 | 57 |
| 6 | 0.820 | 0.673 | 0.819 | 0.838 | 58 |
| 7 | 0.819 | 0.671 | 0.825 | 0.835 | 57 |
| 8 | 0.818 | 0.670 | 0.825 | 0.838 | 57 |
| 9 | 0.818 | 0.669 | 0.817 | 0.837 | 58 |
| 10 | 0.824 | 0.679 | 0.829 | 0.841 | 57 |
| Average | 0.821 | 0.674 | 0.825 | 0.839 | 57 |
| Standard Deviation | 0.002 | 0.004 | 0.004 | 0.002 | 0.9 |

**Table6**

| trial | Pearson Log (r) | $r^2$ | Pearson | Spearman | CPU time (seconds) |
|---|---|---|---|---|---|
| 1 | 0.230 | 0.053 | 0.350 | 0.184 | 18 |
| 2 | 0.224 | 0.050 | 0.365 | 0.185 | 18 |
| 3 | 0.222 | 0.049 | 0.349 | 0.181 | 18 |
| 4 | 0.236 | 0.056 | 0.341 | 0.195 | 19 |
| 5 | 0.202 | 0.041 | 0.112 | 0.165 | 19 |
| 6 | 0.220 | 0.048 | 0.271 | 0.181 | 18 |
| 7 | 0.224 | 0.050 | 0.370 | 0.177 | 18 |
| 8 | 0.219 | 0.048 | 0.170 | 0.174 | 18 |
| 9 | 0.197 | 0.039 | 0.319 | 0.164 | 18 |
| 10 | 0.200 | 0.040 | 0.289 | 0.160 | 18 |
| Average | 0.218 | 0.047 | 0.294 | 0.177 | 18 |
| Standard Deviation | 0.013 | 0.006 | 0.087 | 0.011 | 0.4 |

**Table Captions**

**Table 1.** *Pearson log, Pearson, and Spearman correlations between our method's predictions and experimental probe intensities.*

This table shows the Pearson correlation for log probe intensities and Pearson and Spearman correlations between experimentally measured probed intensities and those predicted by our models. Here we show the results for the 20 TFs with data from both array-types, HK and ME, available.

HK → ME implies models are trained with PBM data from array type HK to predict data obtained from array type ME, and vice versa for ME → HK.

We also show the results of the winner (Annala et al., 2011) of the DREAM5 challenge, teamD. CPU time is obtained on a single processor.

**Table 2.** *Our method's results for the 66 unknown TFs compared with those of the DREAM5 challenge participants.*

Here we show the results obtained after submitting our predictions to the DREAM5 challenge website. Our method ranked better than the winner team in all scoring criteria, and better than all teams in three out of five scores.

**Table 3.** *Pearson of the log (r), $r^2$, Pearson, and Spearman of the synthetic cAMP activity predictions with 10% of sequences used for testing.*

We show the results for ten different trials where we randomly pick 90% of sequences to use for training and 10% for testing. CPU time is obtained on a single processor.

**Table 4.** *Pearson of the log (r), $r^2$, Pearson, and Spearman of the synthetic cAMP activity predictions with 25% of sequences used for testing.*

We show the results for ten different trials where we randomly pick 75% of sequences to use for training and 25% for testing. CPU time is obtained on a single processor.

**Table 5.** *Pearson of the log (r), $r^2$, Pearson, and Spearman of the synthetic cAMP activity predictions with 50% of sequences used for testing*

We show the results for ten different trials where we randomly pick 50% of sequences to use for training and 50% for testing. CPU time is obtained on a single processor.

**Table 6.** *Pearson of the log (r), $r^2$, Pearson, and Spearman of Ifnb activity predictions*

We show the results for ten different trials where we randomly pick 75% of sequences to use for training and 25% for testing. CPU time is obtained on a single processor.

**Figure Captions**

**Figure 1.** *Initial and corrected predictions of log probe intensities for Nr2c1*

a) Histogram of our model's prediction of log probe intensities, $\ln(w_{pred})$, for Nr2c1. The histogram consists of 80 bins. Bins # 2, 64-66, 68-70, and 72-79 are empty.

b) The histogram of $\ln(w_{pred})$ after rectifying the histogram in (a) where points in bin #1 are moved to bin # 2 while points in bins # 67, 71, and 80 are moved to bins # 64, 65, and 66, respectively.

c) Scatter plot of predicted log intensities before (red) and after (blue) rectifying the histogram. True or experimental log intensities, $\ln(w)$, are plotted on the x-axis and the predicted log intensities, $\ln(w_{pred})$, on the y-axis.

**Figure 2.** *Comparing between our predicted motif logos and the JASAPAR sequence logos*

We show the logos obtained from our predictions (right column) for eight different TFs using only the 200 8-mers with the highest KL divergence value. Each is compared with its JASPAR sequence logo on the left.

Note that the JASPAR logo for Irf2 is the reverse complementary sequence.

**Figure 3.** *Goodness of fit variation with number of sequences for Foxo4*

Goodness of fit scores for Foxo4 using different number of sequences (x-axis) in the training set randomly chosen from the HK data set. Each point represents the average value of the Pearson log over 10 trials using different $N$ randomly chosen sequences where $N$ is the number of sequences indicated on the x-axis. The error bars are the standard deviations within the 10 trials.

a) Pearson correlation (y-axis) between predicted log intensities and actual log intensities from the ME data set using different number of sequences (x-axis) in the training set randomly chosen from the HK data set.

b) Spearman correlation (y-axis) between predicted and actual intensities.

c) The area under the receiver operating characteristic (AUROC) for the 35-mer probe intensities.

d) The area under the precision-recall curve (AUPR) for the 35-mer probe intensities.

In (c) and (d) we calculate the AUROC and AUPR values by setting the threshold for bright probes to be 4 standard deviations above the mean of the actual probe intensities. We then rank our predictions and map the receiver operator characteristic (ROC) and precision-recall (PR) spaces.

**Figure 4.** *Goodness of fit variation with number of sequences for Zscan20* Goodness of fit scores for Zscan20 using different number of sequences (x-axis) in the training set. All figures are obtained exactly as described in Figure 3.

a) Pearson correlation between predicted log intensities and actual log intensities for Zscan20.

b) Spearman correlation between predicted and actual intensities for Zscan20

c) AUROC for the 35-mer probe intensities for Zscan20.

d) AUPR for the 35-mer probe intensities for Zscan20.

**Figure 5.** *Motif prediction for CRE*

Our motif prediction for CRE using only the 400 6-mers with the highest KL divergence value.

**Figure 6.** *The CRE sequence*

We draw the CRE sequence used as the reference sequence in (Melnikov et al., 2012), and show the known binding sites, cryptic regions, and the section of the sequence obtained in Figure 5.

**Figure 7.** *Mutual information matrices for CRE*

The mutual information matrices for CRE at $M = 0.1$, $\beta = 1.0$, and using only the 10% highest activity sequences (a), 18% (b), 25% (c), and 50% (d).

## Supporting Information

**Text S1.**

**Table S1.** *Pearson of the log (r), $r^2$, Pearson, and Spearman of the synthetic cAMP activity predictions with 50% of sequences used for testing*

**Figure S1.** *Comparing predicted log intensities obtained by using the highest 200 7-mers and 200 8-mers and those obtained by using the highest 300 7-mers and 300 8-mers*

We show the scatter plots of predicted log intensities for Cebpb (top) and Egr2 (bottom). Experimental log intensities are plotted on the x-axis. On the y-axis, we plot predicted log intensities obtained by using the highest 200 7-mers and 200 8-mers (red) and those obtained by using the highest 300 7-mers and 300 8-mers (blue).

**Figure S2.** *Comparing predicted log intensities at different M values*

We show the scatter plots of predicted log intensities for Cebpb (top) and Egr2 (bottom) using the highest 200 7-mers and 200 8-mers. On the y-axis we plot predicted log intensities at $M = 0.1$ (red), $M = 0.09$ (blue), and $M = 0.08$ (green).

**Figure S3.** *Predicted motif logos using both 7-mers and 8-mers*

We show the logos obtained from our predictions (right column) for eight different TFs using the 200 7-mers and 200 8-mers with the highest KL divergence value. Each is compared with its JASPAR sequence logo on the left.

**Figure S4.** *Predicted motif logos using 6-mers, 7-mers, and 8-mers*

We show the logos obtained from our predictions (right column) for eight different TFs using the 1000 6-mers, 200 7-mers, and 200 8-mers with the highest KL divergence value. Each is compared with its JASPAR sequence logo on the left.

**Figure S5.** *Average log activities over sequences with the same number of mutations*

The histogram shows the fraction of sequences having a certain number of mutations. It is the same for both CRE and Ifnb. We also plot the average of log activities, $\ln(w)$, over sequences with a set number of mutations vs the corresponding number of mutations (x-axis) for both CRE (red) and Ifnb (blue).

**Figure S6.** *Motif prediction for the uninduced CRE*

Our motif prediction for the uninduced CRE using only 400 8-mers with the highest KL divergence.

The data for the uninduced CRE are also obtained from (Melnikov et al., 2012). We follow the same procedures we used to obtain a motif prediction of the induced CRE in Figure 5, except that here we use 8-mers instead of 6-mers as the former lead to better performance. We find that positions 9 to 20 of the predicted motif correspond to positions 7 to 18 of the reference sequence shown in Figure 6. These positions include part of the first cryptic site and the first binding site, CREB(1), which is consistent with the results obtained by (Melnikov et al., 2012) showing that CREB(1) has the largest contribution in enhancing the activity of the uninduced CRE. The other positions of the motif cannot be mapped, however, to any part of the reference sequence.

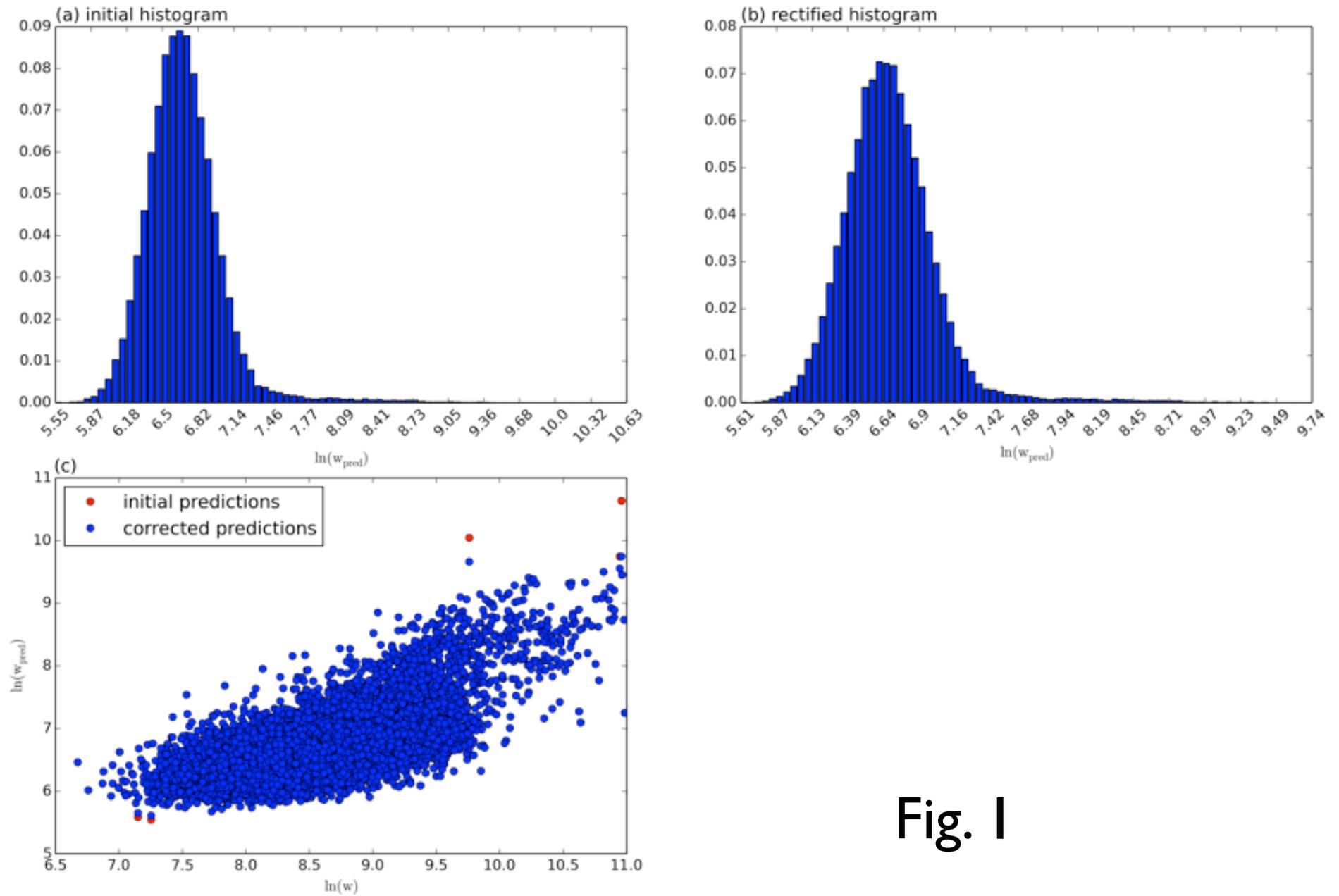

Fig. 1

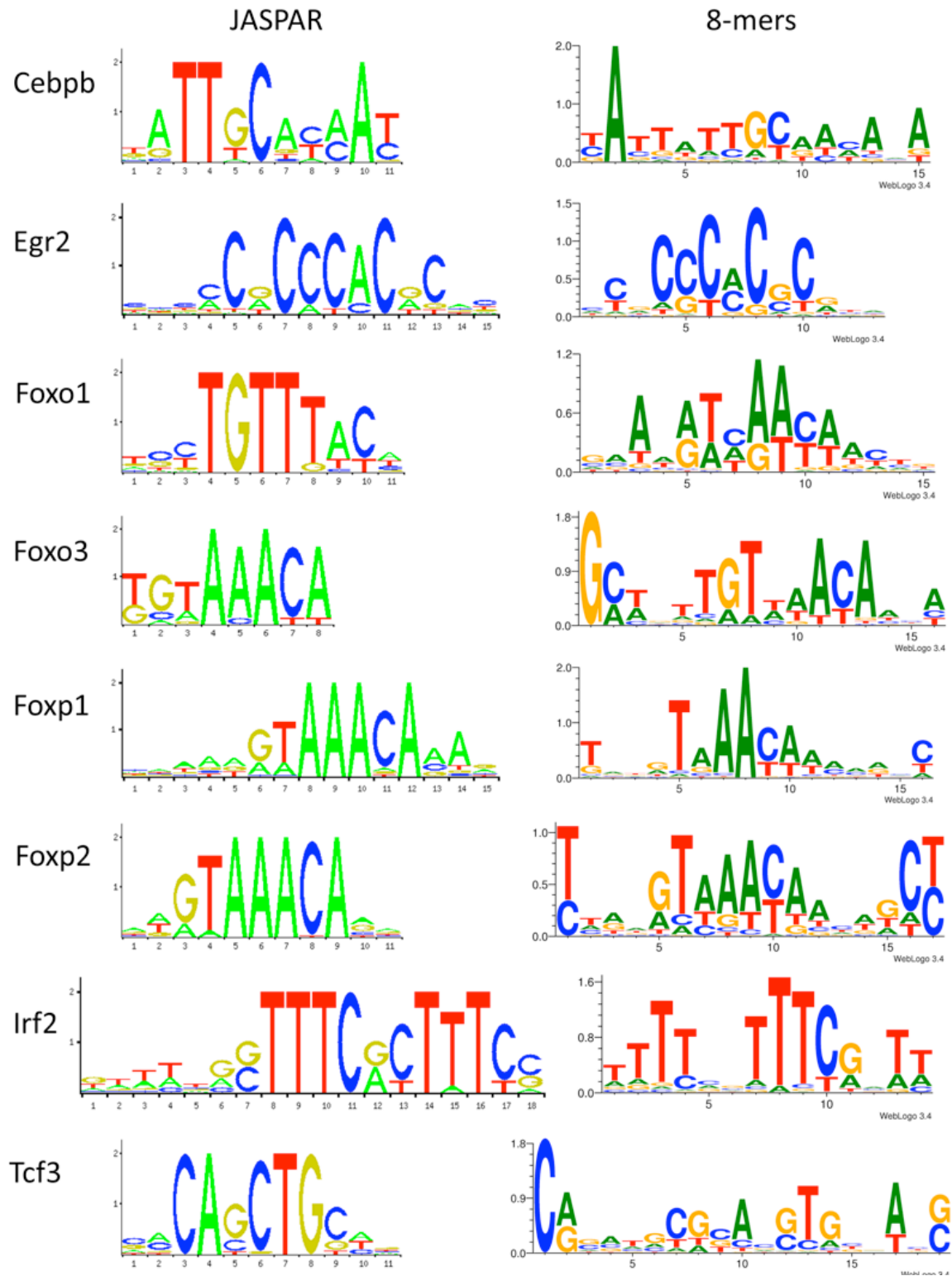

Fig. 2

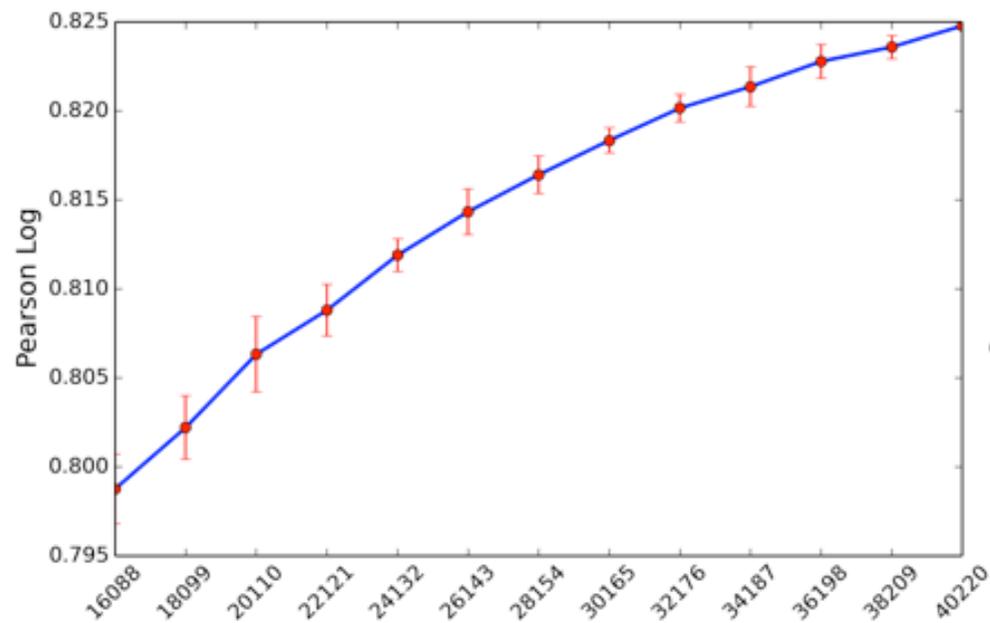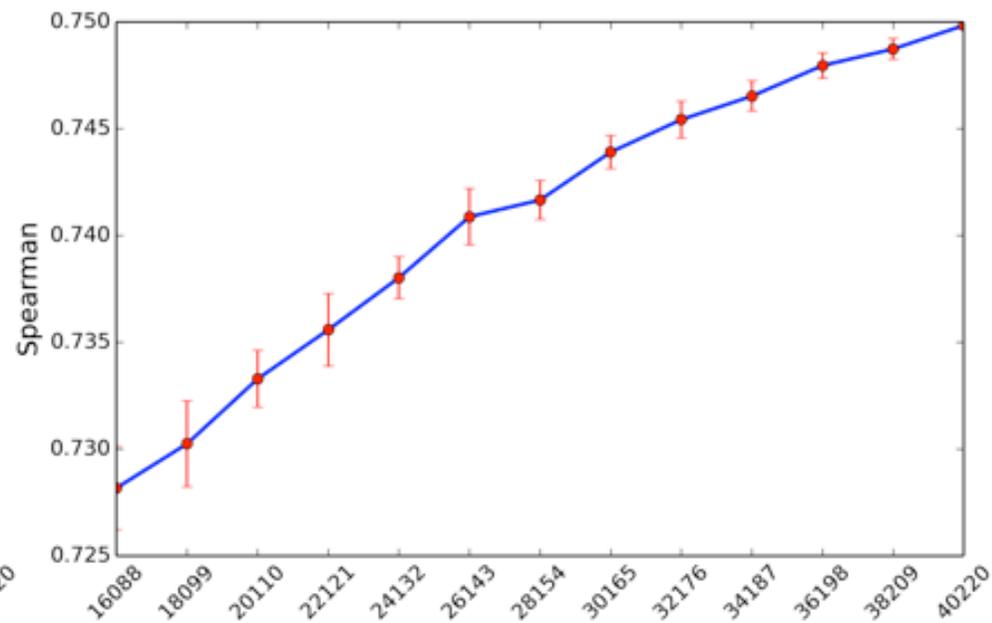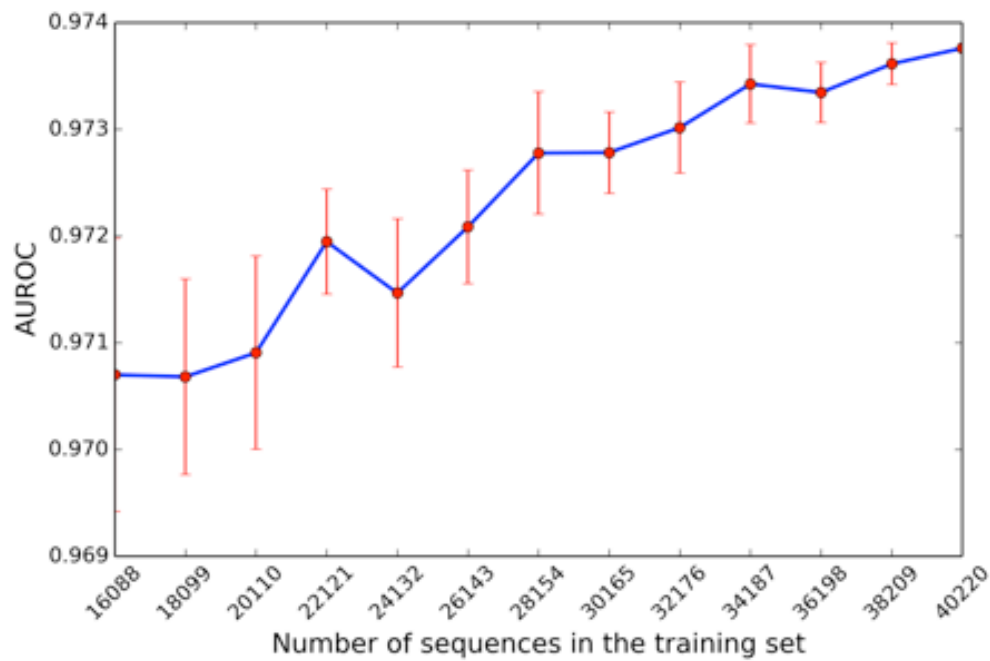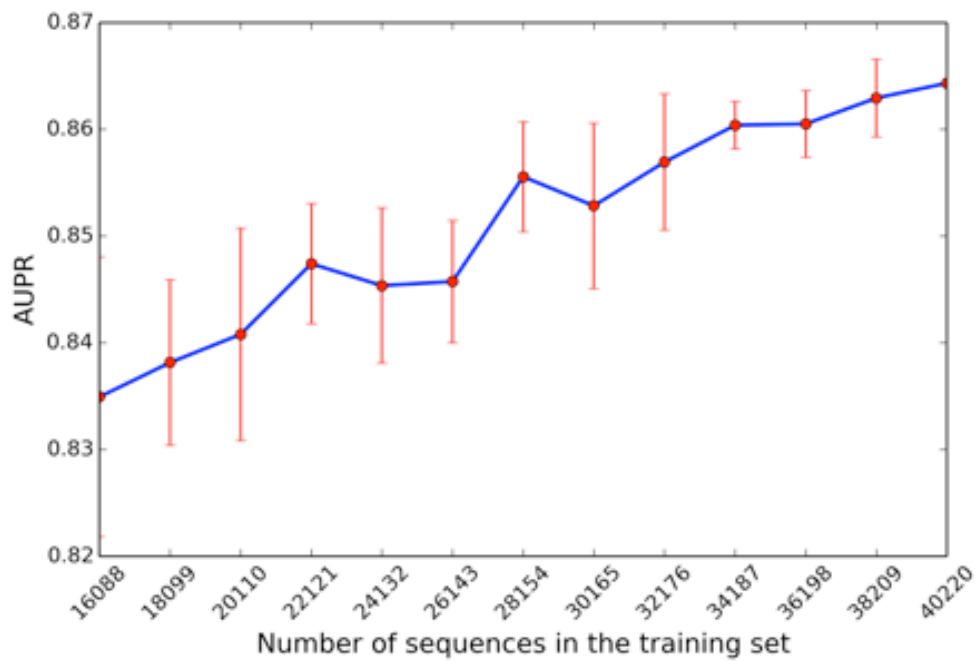

Fig. 3

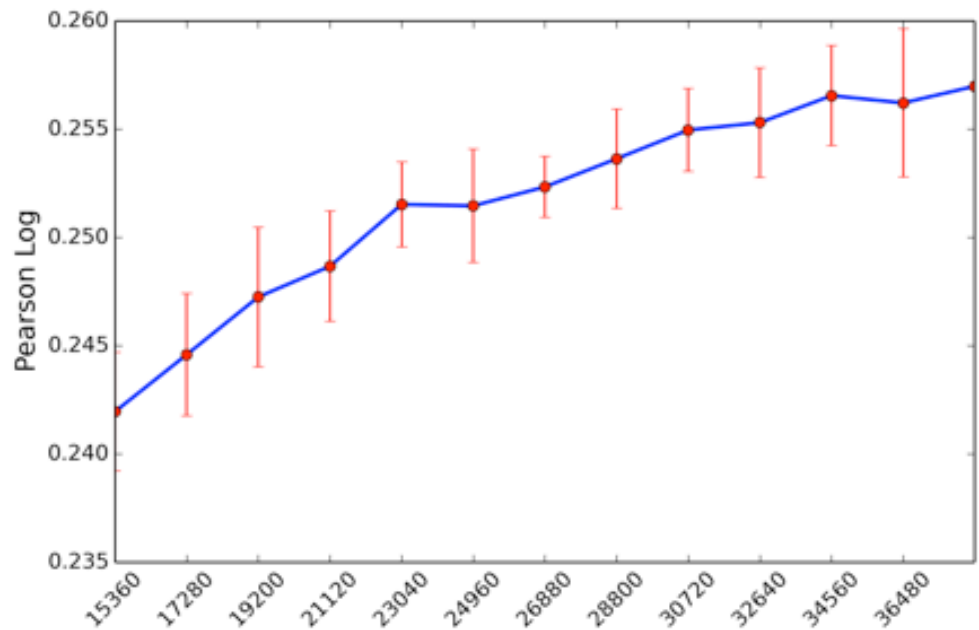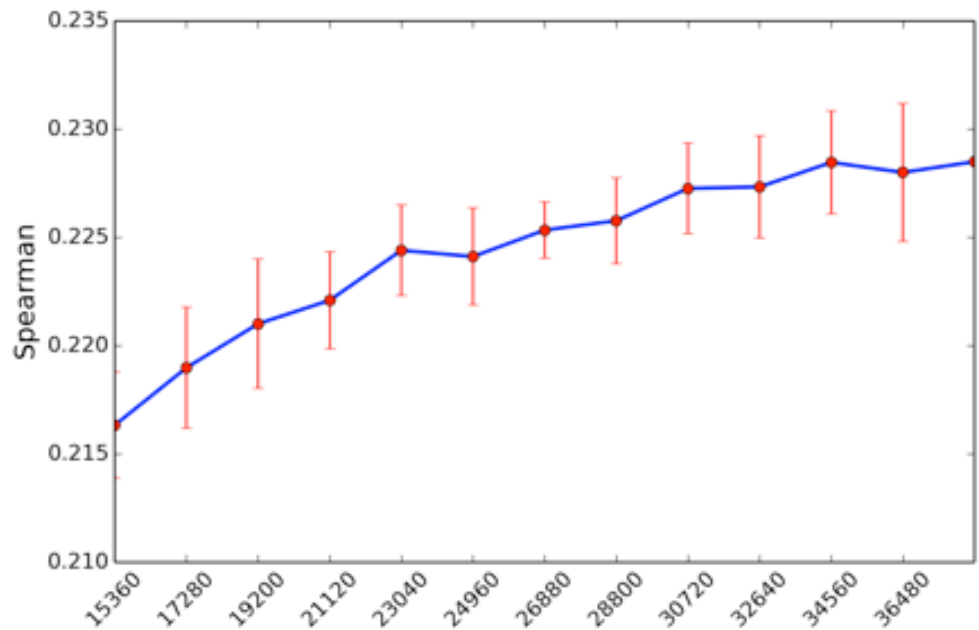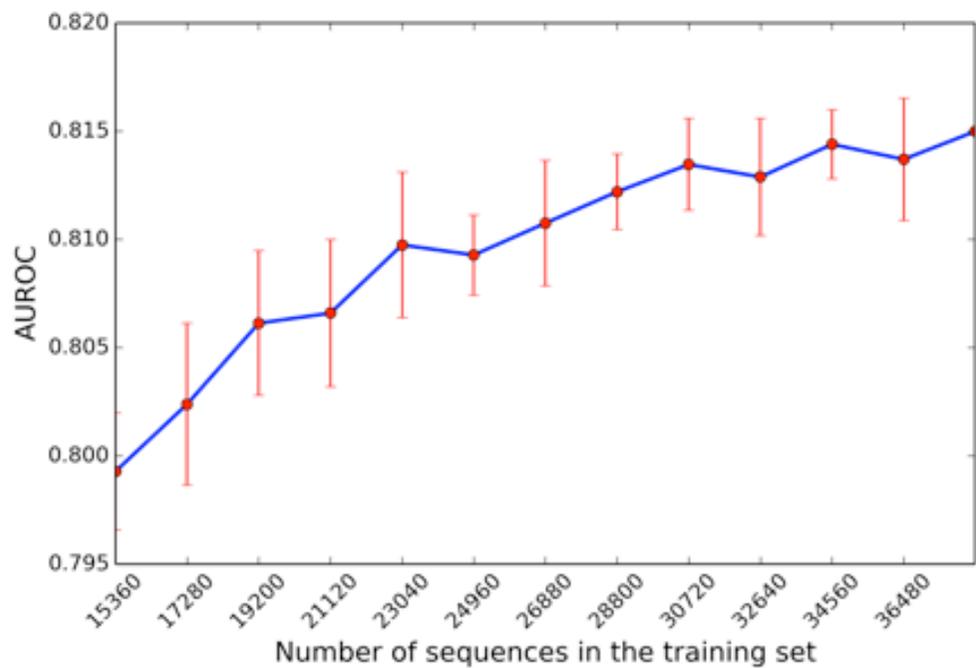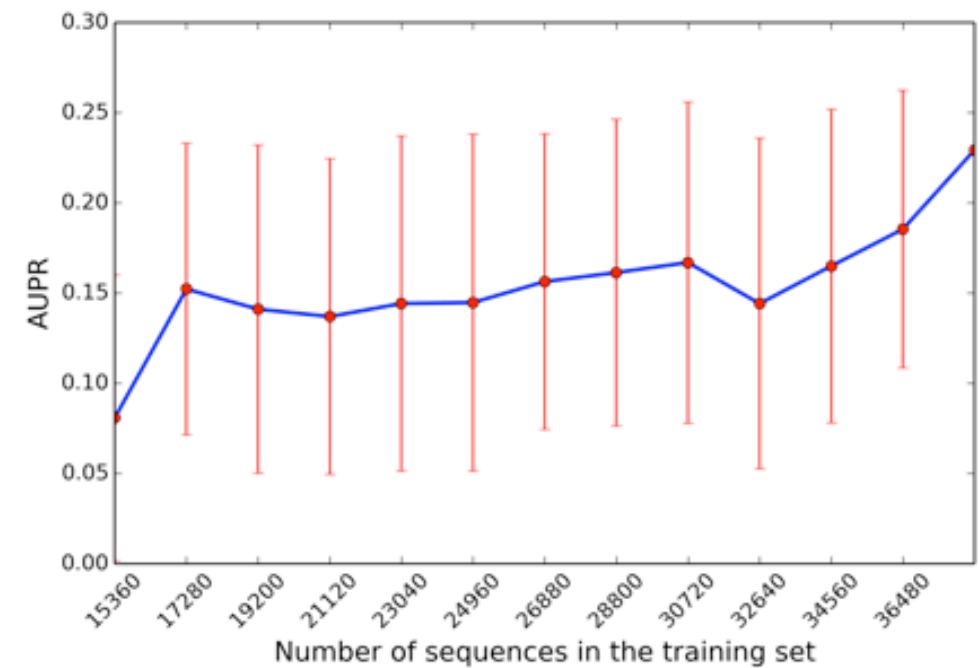

Fig. 4

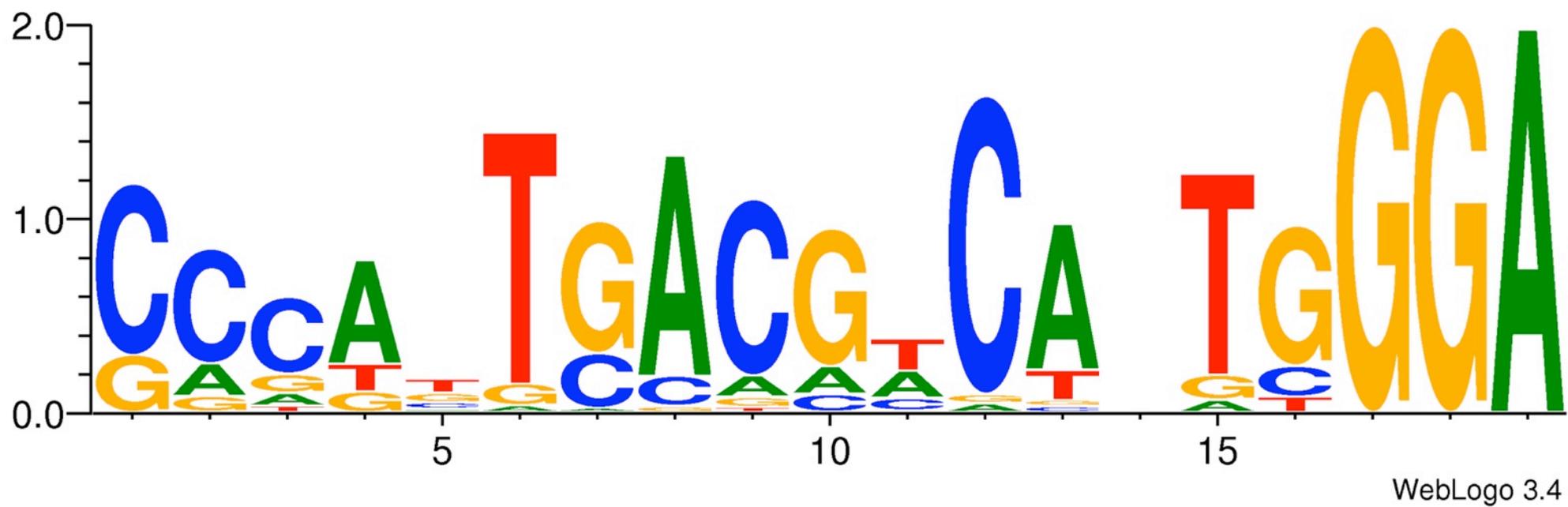

Fig. 5

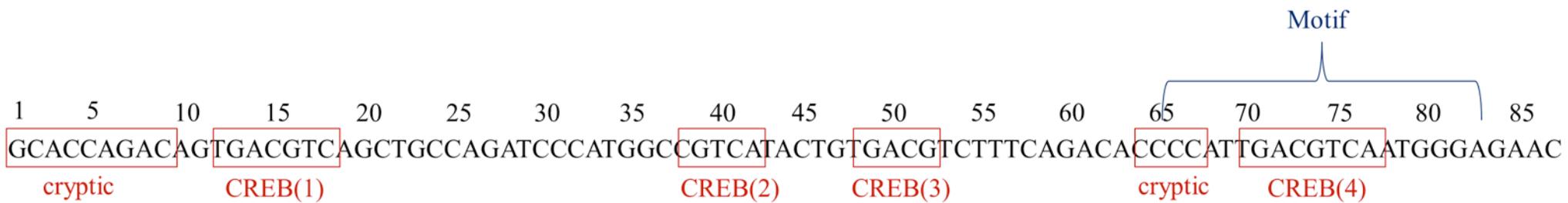

Fig. 6

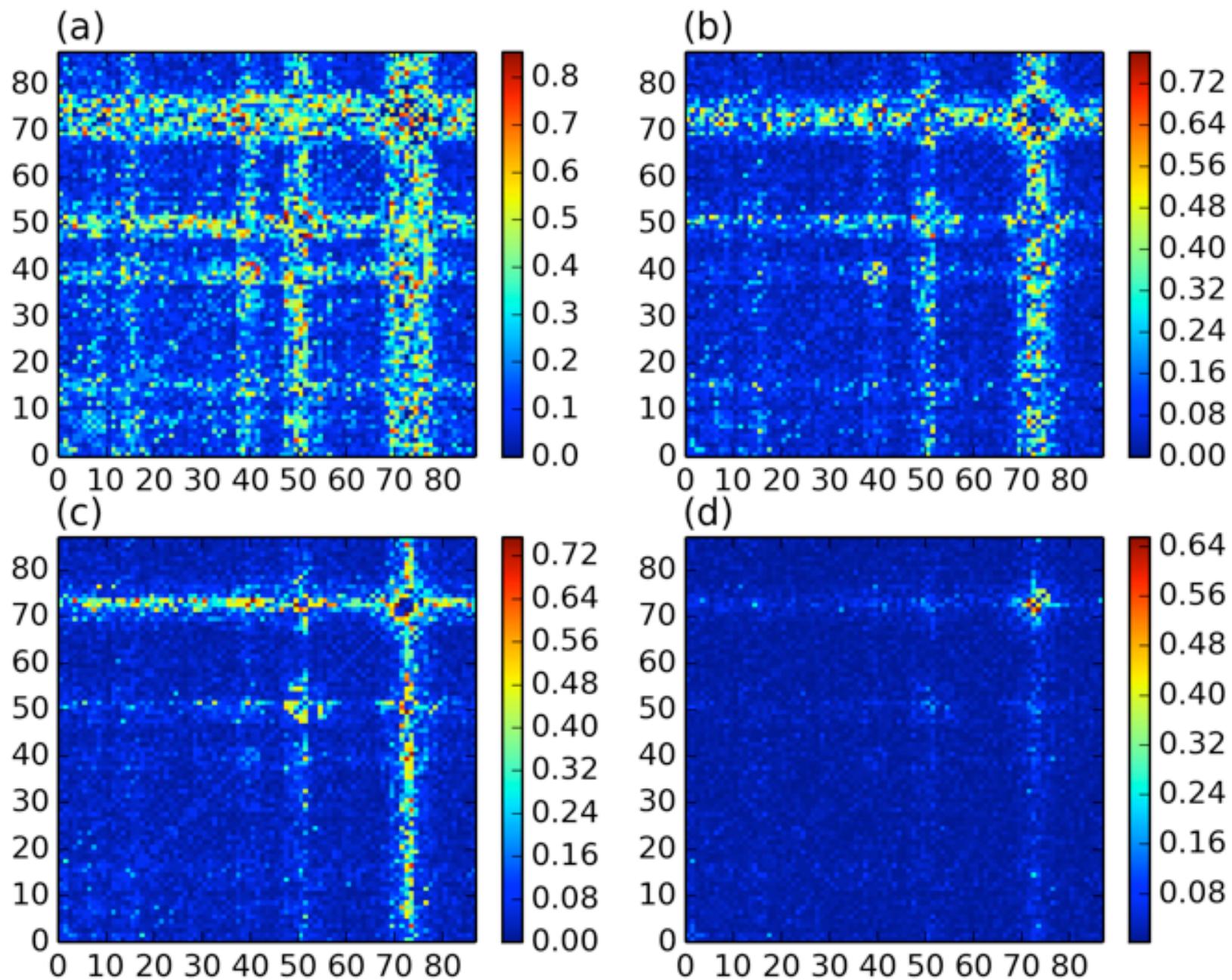

Fig. 7

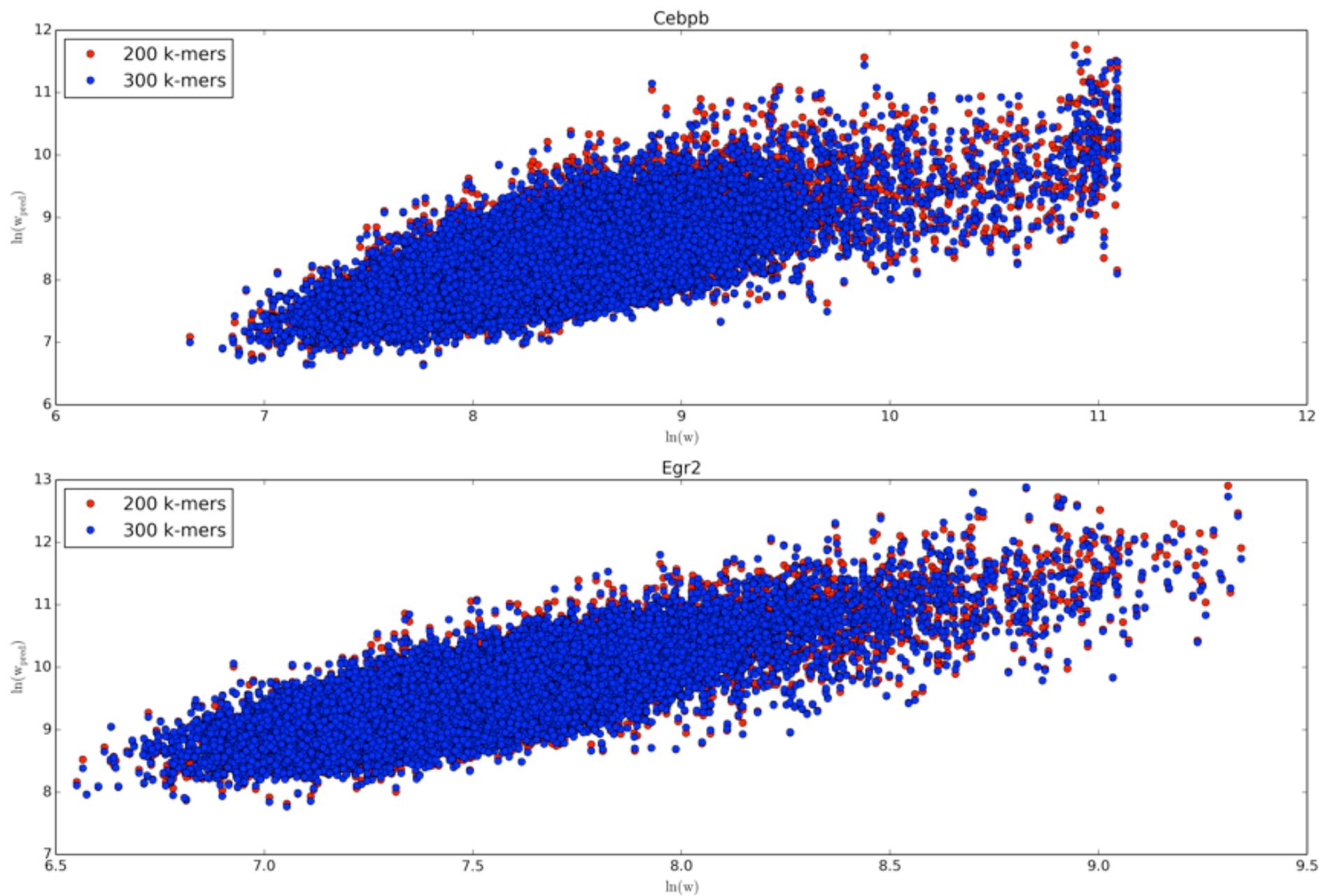

Fig. S1

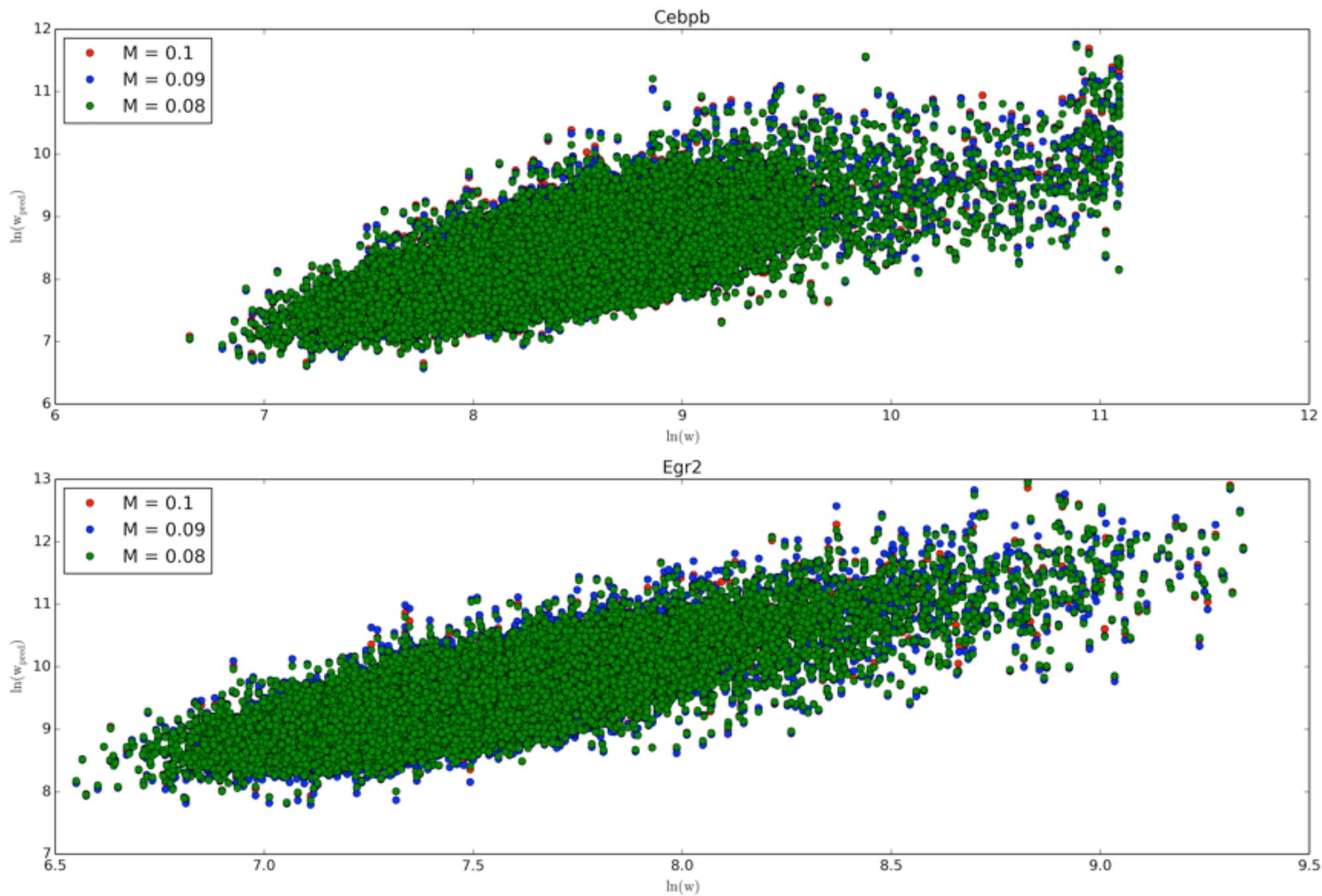

Fig. S2

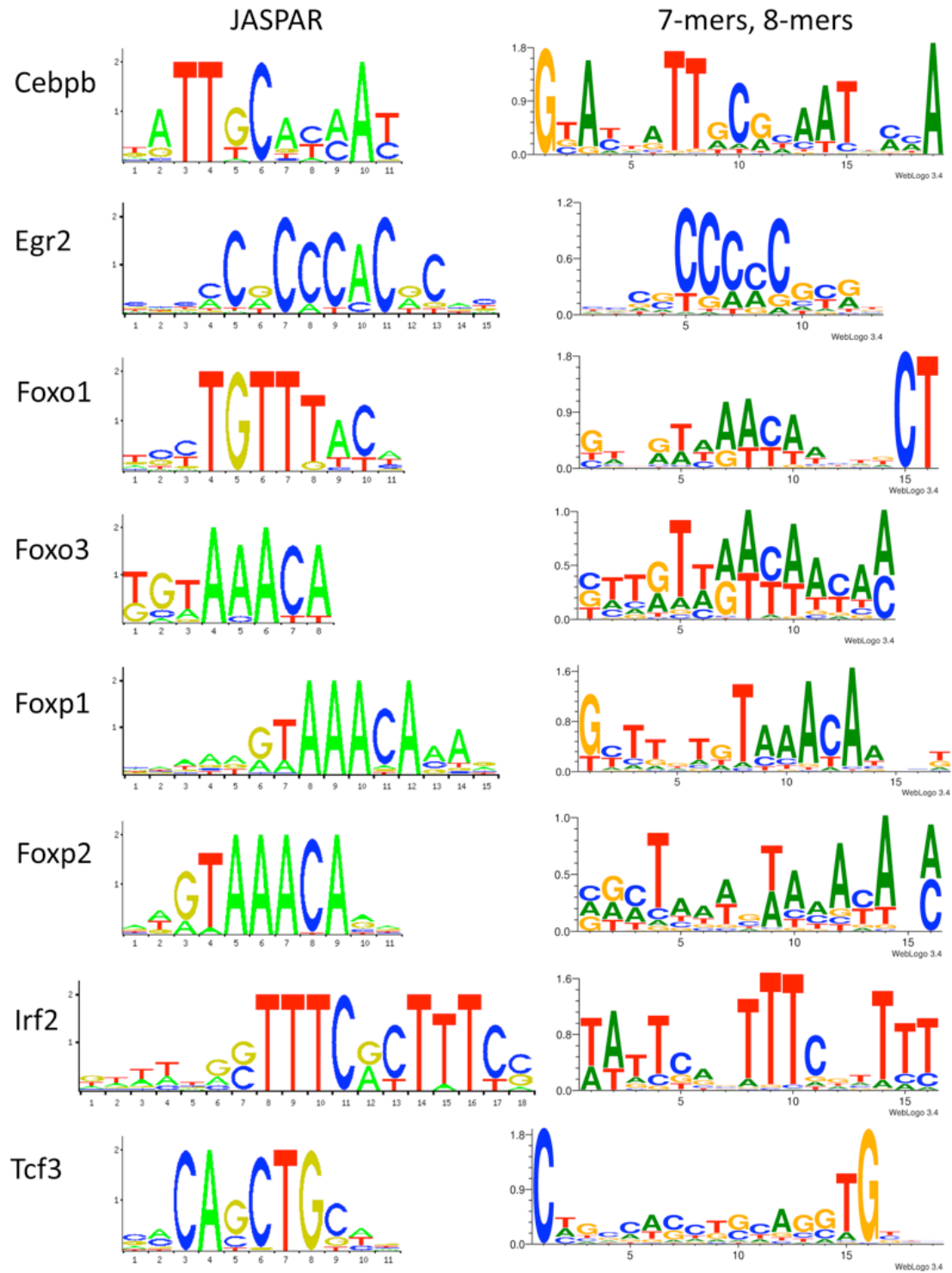

Fig. S3

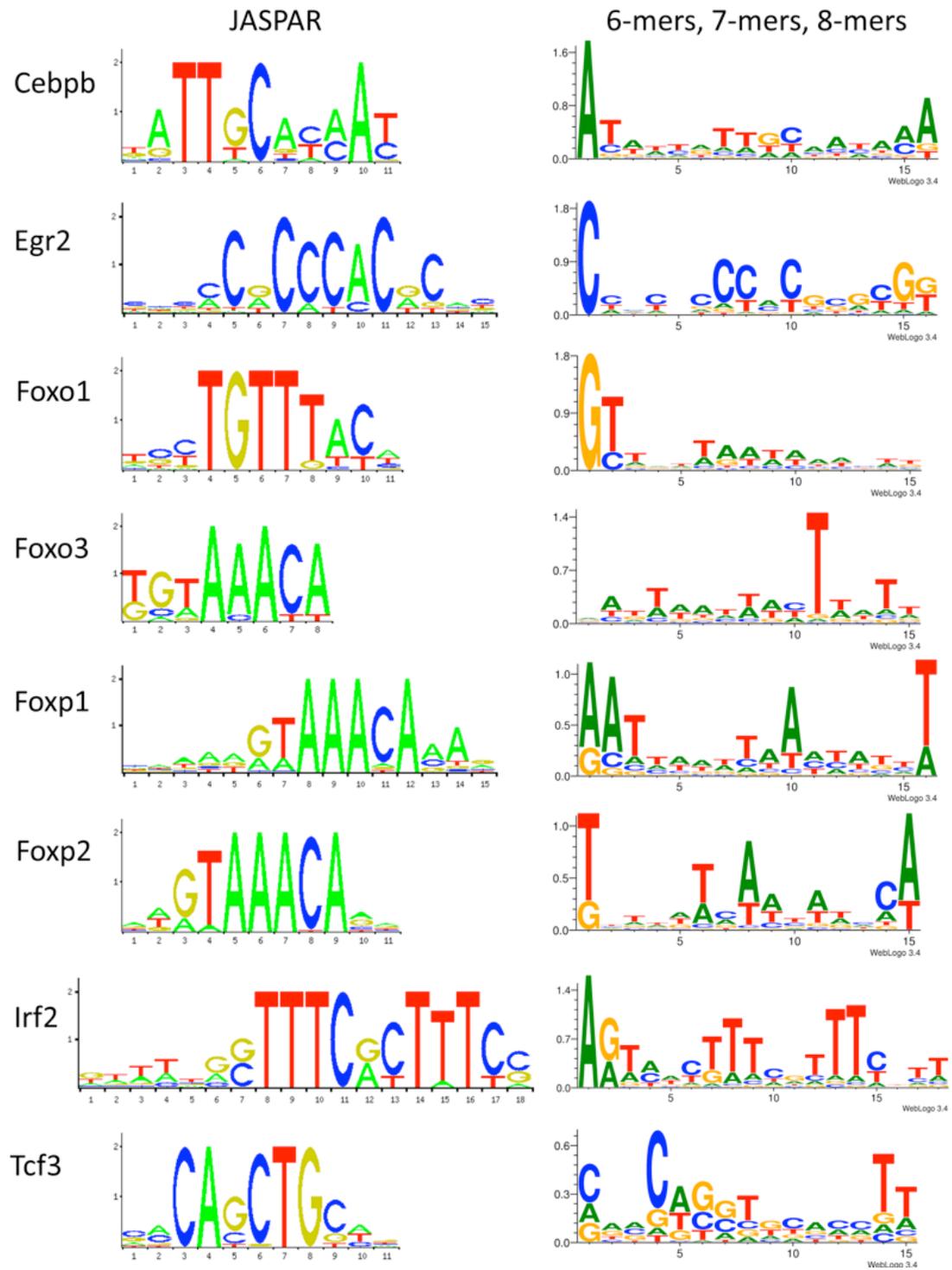

Fig. S4

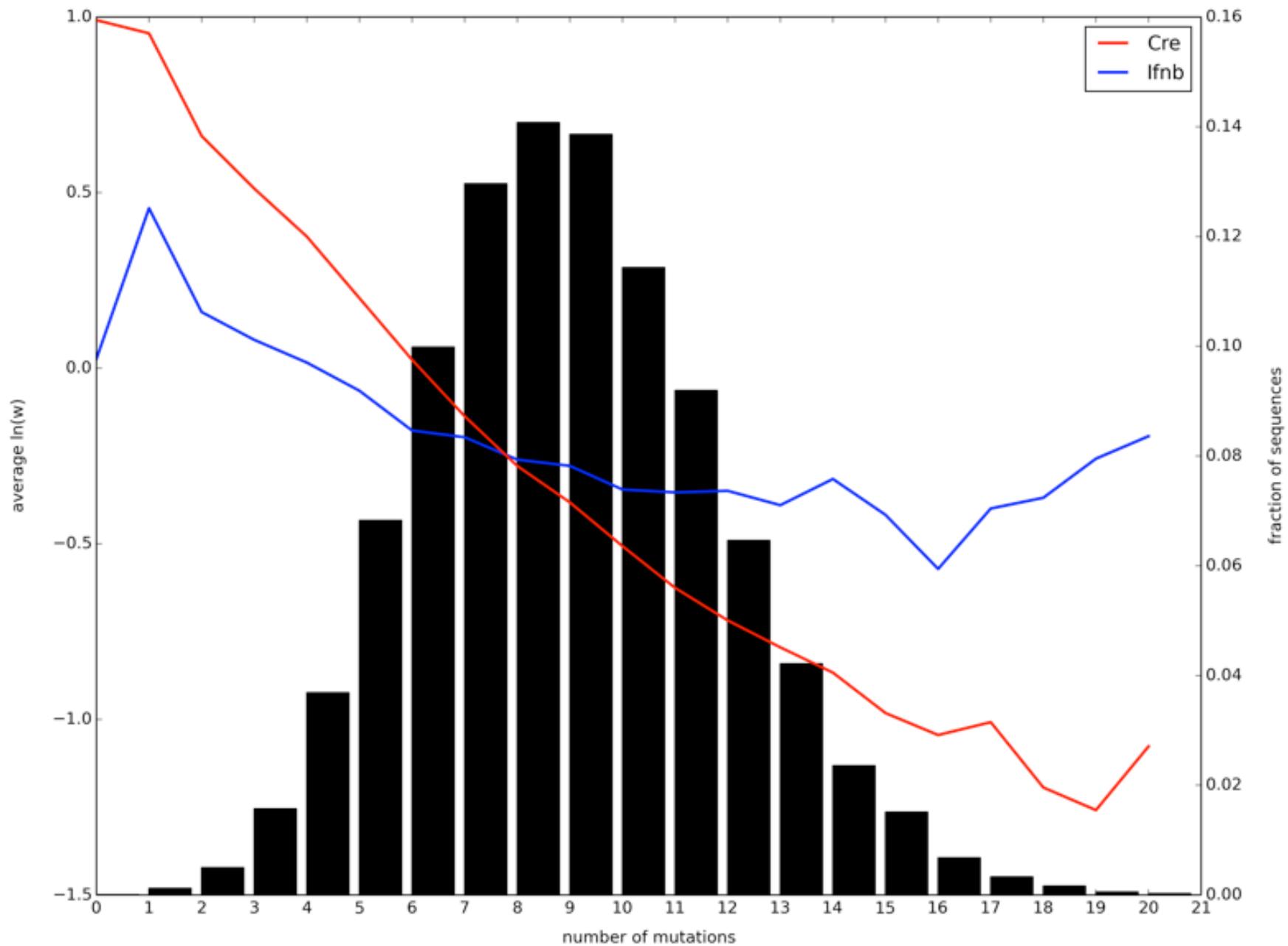

Fig. S5

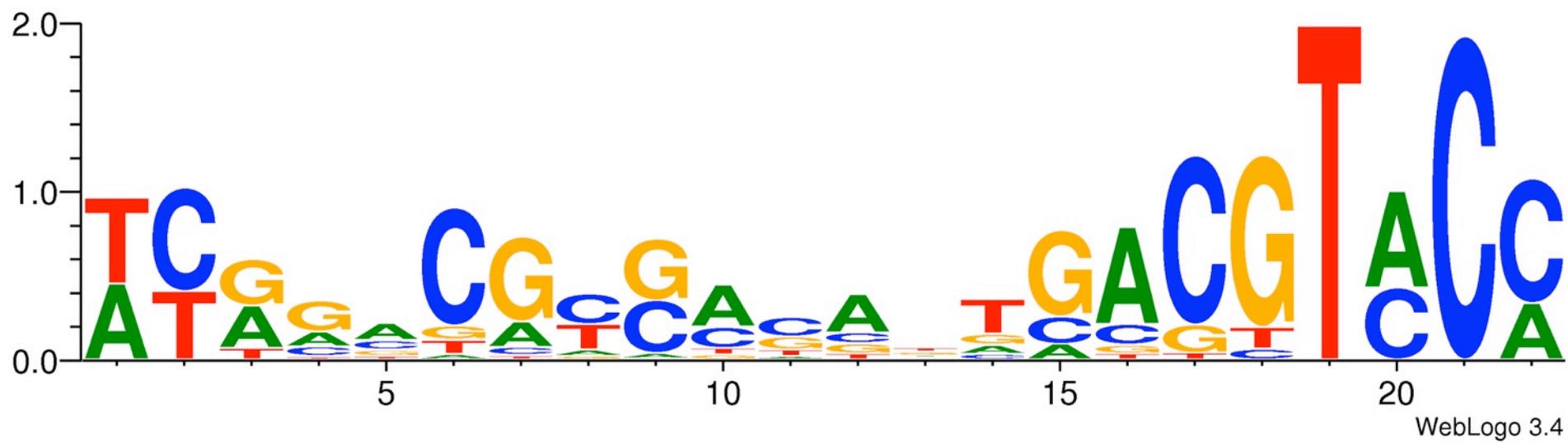

Fig. S6

**6. Supplementary Material for on-line publication only**

[Click here to download 6. Supplementary Material for on-line publication only: QSAM_TextS1_JTB_vp.docx](#)